\documentclass[11pt]{article}
\usepackage{jheppub}

\usepackage{graphicx}
\usepackage{genyoungtabtikz}
\usepackage{tikz,varwidth}
\usetikzlibrary{shapes,snakes}
\usetikzlibrary{arrows, decorations.markings}
\usetikzlibrary{matrix,decorations.pathreplacing,calc}
\usetikzlibrary{decorations.text,calc,arrows.meta}

\def\be{\begin{equation}}
\def\ee{\end{equation}}
\def\ba{\begin{eqnarray}}
\def\ea{\end{eqnarray}}

\def\e{\epsilon}

\def\IR{\relax{\rm I\kern-.18em R}}

\def\inv{^{\raise.0ex\hbox{${\scriptscriptstyle -}$}\kern-.05em 1}}

\title{BMN Vacua, Superstars and Non-Abelian T-duality}

\author{Yolanda Lozano$^{1}$,}
\author{Carlos N\'u\~nez$^2$}
\author{and Salom\'on Zacar\'{\i}as$^2$}

\affiliation{$^1$ Department of Physics, University of Oviedo, Avda. Calvo Sotelo 18, 33007 Oviedo, Spain}
\affiliation{$^2$ Department of Physics, Swansea University, Swansea SA2 8PP, United Kingdom}

\emailAdd{ylozano@uniovi.es} 
\emailAdd{c.nunez@swansea.ac.uk} 
\emailAdd{salomon.zacarias@swansea.ac.uk} 

\abstract{Acting with non-Abelian T-duality on the $S^3$  inside the $AdS_5$ subspace of $AdS_5\times S^5$ with $N$ units of flux, we generate a new half-BPS solution with $SU(2|4)$ symmetry that belongs to  the Lin-Lunin-Maldacena class of geometries. The analysis of the asymptotics, quantised charges and probe branes in this geometry suggests an interpretation as the gravity dual to the Berenstein-Maldacena-Nastase Plane Wave Matrix Model, in a particular vacuum associated to a partition of $N$, in which the multiplicity of each $SU(2)$ irreducible representation is equal to its dimension. This vacuum is interpreted in M-theory in terms of giant gravitons backreacting in the maximally supersymmetric pp-wave geometry. Consistently with this, we show that the non-Abelian T-dual solution exactly agrees with the Penrose limit of the superstar solution in $AdS_7\times S^4$. This suggests an interesting global completion of the non-Abelian T-dual solution in terms of an M5-brane geometry.}


\dedicated{Dedicated to the memory of  Haim Goldberg.}

\begin{document}
\def\Tr{{\textrm{Tr}}}
\def\be{\begin{equation}}
\def\e{\end{equation}}
\def\bea{\begin{equation*}}
\def\ea{\end{equation*}}
\def\la{\label}
\def\bu{\bullet}

\maketitle 
\newpage
\section{Introduction}
Among the different dualities in Quantum Field Theory \cite{Polchinski:2014mva} and in String Theory \cite{Witten:1995ex}, non-Abelian T-duality is probably 
among the least understood ones. The idea of extending Buscher's procedure \cite{Buscher:1987sk,Rocek:1991ps}
 to the case of non-Abelian isometries, proposed by de la Ossa and Quevedo in \cite{delaOssa:1992vci},  motivated
thorough studies. Some  stringy features (mostly the sigma-model aspects of the transformation) were spelled out in different publications during the mid-nineties---see for example \cite{Alvarez:1993qi,Giveon:1993ai}. These papers made the point that global aspects of the duality in \cite{delaOssa:1992vci} are ill-understood. Indeed, for world-sheets  of genus bigger than zero, it is presently unclear how to make the procedure consistent. It was then suggested that the non-Abelian generalisation of the Buscher procedure and the  transformations in \cite{delaOssa:1992vci} were actually {\it not} a duality, but a transformation between different world-sheet conformal field theories
\cite{Giveon:1993ai}. All these discussions concerned only the NS-NS sector of the string theory. These studies  were followed by a period of  low-activity in the subject.

It was the 2010 paper of Sfetsos and Thompson \cite{Sfetsos:2010uq}, that gave a prescription to consistently transform the Ramond-Ramond fields, which reignited the interest in the field. Indeed, the synergy between non-Abelian T-duality and the AdS/CFT correspondence \cite{Maldacena:1997re}, lead to various  lines of research:
\begin{itemize}
\item{The use of non-Abelian T-duality as a solution generating technique resulted in new  backgrounds that in some cases avoid previously studied classifications \cite{variosa1,Itsios:2013wd,variosa2,Lozano:2014ata,variosa3,Macpherson:2014eza,variosa4,Macpherson:2015tka,variosa5}.}
\item{Using the newly generated backgrounds, new quantum field theories at strong coupling have been defined via AdS/CFT and through explicit semiclassical calculations 
 \cite{Itsios:2013wd,Lozano:2013oma,variosb1,Macpherson:2015tka, Bea:2015fja,Lozano:2016kum,Lozano:2016wrs,INZ}.}
\item{Defining a string sigma-model in these new backgrounds new relations with integrable models have been worked out \cite{Hoare:2016wsk}. }
\end{itemize}
Some of the papers quoted above combine more than one of these lines of research.

Since it happens that {\it some} of the backgrounds generated by non-Abelian T-duality are singular, the study of the properties of the  field theory dual to those singular spaces, is by definition incomplete. Indeed, in such cases one cannot  use the string background to learn about non-perturbative aspects of the dual field theory. This is the case for some paradigmatic examples, like the non-Abelian T-dual of $AdS_5\times S^5$---see for example \cite{Sfetsos:2010uq,Macpherson:2015tka,Macpherson:2014eza}. For such examples (that characteristically have large isometry group and number of SUSYs), the idea proposed in  \cite{Lozano:2016kum} is to use the field theory (that in cases with large global symmetry groups is sharply determined) to {\it amend} or {\it complete} the background, curing its singular behaviour. In other words, the quantum field theory {\it dictates} the geometry.
This idea was made explicit in the  papers
 \cite{Lozano:2016kum} and \cite{Lozano:2016wrs}, where an ${\cal N}=2$ super-conformal field theory in four dimensions and an ${\cal N}=4$ super-conformal 
  field theory in three dimensions, respectively, dictate the way to resolve the singular behaviour of the backgrounds obtained using non-Abelian T-duality.

In this paper, a particularly interesting realisation of these ideas is studied. The general outcome of the papers \cite{Lozano:2016kum} and \cite{Lozano:2016wrs}, that backgrounds obtained via non-Abelian T-duality are a small {\it patch} of a more general geometry, will work-out explicitly in our example. The precise way in which such patch will be selected is via a Penrose limit. Indeed, as we will see, the new non-Abelian T-dual background studied in this paper is the Penrose (pp-wave) limit of the so-called Superstar solution \cite{Leblond:2001gn}.  This is the first explicit realisation of a loose but intuitive conjecture that can be found scattered in different papers---see for example \cite{Polychronakos:2010hd,Sfetsos:2010uq}.

The precise set-up in which we are able to find this explicit realisation is a $R_t\times SO(3)\times SO(6)$ symmetric solution that we construct acting with non-Abelian T-duality on the $AdS_5$ subspace of $AdS_5\times S^5$. This solution is dual to a particular vacuum of the large class of 1/2 BPS vacua of the Berenstein, Maldacena, Nastase (BMN) Plane Wave Matrix Model  \cite{Berenstein:2002jq}.  These vacua are interpreted in the dual supergravity solution as D0-branes polarised into D2-branes at weak coupling, or into NS five branes at strong coupling, very similarly to the ${\mathcal N}=1^*$ Super-Yang-Mills  vacua structure \cite{Polchinski:2000uf}. Using that the BMN matrix model is a massive deformation of the BFSS matrix model \cite{Banks:1996vh}, Lin applied the method in \cite{Polchinski:2000uf} to derive a set of equations that the gravity dual solutions to BMN vacua should satisfy \cite{Lin:2004kw}. The paper of Lin and Maldacena \cite{Lin:2005nh} finally identified the precise electrostatic problem that must be solved to construct explicitly these solutions within the formalism of \cite{Lin:2004nb}. 

Let us be more detailed about the contents and plan of this paper:
\begin{itemize}
\item We start our paper by going through the main results of some of the previously mentioned  papers, of relevance for our work. These are summarised in Section \ref{pwmm}.

\item{In Section \ref{sectionNATD}, we construct the new solution with $SU(2|4)$ isometry supergroup (this is $R_t\times SO(3)\times SO(6)$ and 16 SUSYs). We study its asymptotics, show that it satisfies the set of equations posed by Lin in \cite{Lin:2004kw} and identify the precise way in which it enters
in the classification of  \cite{Lin:2005nh}. We calculate conserved charges and interesting probe branes that characterise our new background and allow to interpret it as dual to a particular BMN vacuum. }
\item{In Section \ref{QFT},  we present a ($0+1$)-dimensional quantum field theory that captures the dynamics of the non-Abelian T-dual solution and dictates a {\it completion} of the background. This allows to interpret our solution as a patch of a more general geometry dual to a relevant deformation of the BMN Plane-Wave Matrix Model  \cite{Berenstein:2002jq}.}
\item{The material in Section \ref{penrosesection} contains the proposal of a different completion to our non-Abelian T-dual, as a Penrose limit of the Superstar geometry on $AdS_7\times S^4$. Through this completion, the quantum field theory dual to the non-Abelian T-dual background is the six dimensional (0,2) conformal theory deformed by relevant operators.}
\item{Using an  analytic continuation, we connect our new background with that found by Sfetsos and Thompson in \cite{Sfetsos:2010uq}. This is presented in Section \ref{connectionst}. }
\end{itemize}
Finally, some conclusions  and possible future research directions are collected in Section \ref{conclusiones}, which  closes our paper.

%
\section{The BMN Matrix Model and its gravity dual}\label{pwmm}
In this section, we summarise different results pertaining to the Plane Wave Matrix Model (PWMM) introduced by Berenstein, Maldacena and Nastase in \cite{Berenstein:2002jq}. These results have been derived in a  large collection of papers, of which we single out  \cite{Lin:2004kw}-\cite{Donos:2010va}. The aim of this section is to organise them in a way  useful for our treatment in Section \ref{sectionNATD}.

We start by considering ${\cal N}=4$ SuperYang-Mills (SYM) defined on $R_t\times S^3$. The field theory has global bosonic symmetries $SO(2,4)\times SO(6)$ and preserves 32 SUSYs. Since the CFT is defined on a three-sphere, it is  natural to distinguish inside $SO(2,4)$ a subgroup $SO(4)\sim SU(2)\times \hat{SU}(2)$. This field theory has a single vacuum state and is dual to the Type IIB string theory on $AdS_5\times S^5$, with the $AdS_5$
written in global coordinates, making explicit the $SO(4)$ isometry of the gravity solution.

We consider  modding out this field theory by subgroups of $\hat{SU}(2)$. By this, we mean erasing  or truncating-away from the Lagrangian of ${\cal N}=4$ SYM on $R_t\times S^3$ all the fields that transform (are not singlets) under the given subgroup.
The subgroups we can choose are $Z_k$, $U(1)$ and the whole $\hat{SU}(2)$. The field theories that result after the modding out are ${\cal N}=4 $ SYM on $R_t\times S^3/Z_k$, ${\cal N}=8$ SYM on $R_t\times S^2$ and the BMN Plane Wave Matrix Model, respectively.
All these quantum field theories  have global symmetries $SO(6)\times SO(3)$ and preserve 16 SUSYs. The three theories have a very interesting vacuum structure with a mass gap. 

In the following and for the purposes of this paper, we will focus on the Plane Wave Matrix Model.
\subsection{The BMN Plane Wave Matrix Model}
As discussed above, this quantum mechanical model is obtained by modding out by $\hat{SU}(2)$ the CFT with 32 SUSYs defined on $R_t\times S^3$. After a suitable rescaling of fields, the Hamiltonian reads \cite{Berenstein:2002jq},
\begin{equation}
H=Tr\Big[  \frac{P_A^2}{2} +\frac{1}{2} ( \frac{X_i^2}{9}  +\frac{X_a^2}{36}) +\frac{i g}{3}\epsilon^{ijk}X_iX_jX_k -\frac{g^2}{4}[X_A,X_B]^2+{\rm fermions}   \Big].
\label{hamiltonianpwmm}
\end{equation}
Here the labels take values $A=1,....,9$; $i=1,2,3$ and  $a=4,5,6,7,8,9$. The scalar fields $X_A$, the conjugate momenta $P_A$ and the omitted 16-component fermions are hermitian $N\times N$ matrices. The parameters characterising the PWMM are $g$, that is continuous, and $ N$, that is discrete. The model has a gauge symmetry that requires observable operators $M$, to transform as
$M\to U^{-1}M U$.  The rotation among the  six $X_a$ fields realises the $SO(6)$ global symmetry and the rotation among the three $X_i$ does the same for the $SO(3)$ symmetry. The model preserves 16 SUSYs. All these global symmetries are compactly expressed by the supergroup $SU(2|4)$. Notice that the Hamiltonian in eq.(\ref{hamiltonianpwmm}) is a massive deformation of the dynamics of D0-branes. Short wave-length excitations are then described by the BFSS matrix model \cite{Banks:1996vh}.

The classical vacua of the PWMM are  found by choosing $X_a=0$. The remaining F-term equation is,
\begin{equation}
\frac{1}{3g} X_i=\frac{i}{2} \epsilon_{ijk}[X_j,X_k].\nonumber
\end{equation}
Its solution is  $3 g X_i=J_i$, being $J_i$ an $N$-dimensional representation of $SU(2)$. In general, the vacuum state of the PWMM is then in one-to-one correspondence with partitions of $N$. We can have $n_k$ copies of the $k$-dimensional representation of $SU(2)$---of dimension $k=(2 j_k +1)$,
\begin{equation}
\label{breaking}
N=\sum_k^T n_k (2 j_k+1).
\end{equation}
The vacuum then breaks spontaneously  the gauge symmetry according to, 
\begin{equation}
U(N)\to U(n_1)\times U(n_2)\times....\times U(n_T). 
\end{equation} 
It is worth stressing that all these vacua are uplifted to the same vacuum in ${\cal N}=4$ SYM, since they give rise to zero energy configurations that differ from the ordinary vacuum by large gauge transformations. 

It is interesting that the BMN action can also be obtained by quantising the supermembrane action or the system of $N$ D0-branes  \cite{Dasgupta:2002hx}, in the maximally supersymmetric 11d pp-wave background. In the last description the F-term in the vacuum equations arises as a Myers dielectric term in the action of the $N$ D0-branes. The interpretation of the non-trivial vacua is then in terms of $N$ D0-branes expanded into D2-branes with the topology of fuzzy spheres.

This vacua structure is very similar to that of the  
${\cal N}=1^*$ SYM theory studied by Polchinski and Strassler \cite{Polchinski:2000uf}, which also contains fuzzy sphere vacua, associated in this case to D3-branes polarising into D5.  In the same vein, the D0 branes represented by eq.(\ref{hamiltonianpwmm}) can also polarise into NS5-branes at strong coupling \cite{Maldacena:2002rb}.
 
Let us now discuss the Type IIA backgrounds that are dual to the different vacua of the BMN matrix model.

\subsection{The Type IIA String Duals}\label{method}
In the interest of space, we will not summarise here the results of the remarkable paper 
by Lin-Lunin-Maldacena  \cite{Lin:2004nb}  (LLM), that started the classification of backgrounds preserving 16 SUSYs
with a given set of isometries. The formalism of LLM has applications to a wide class of examples, being the PWMM one of them. 

It is clear that backgrounds preserving $SO(6)\times SO(3)\times R_t$ in M-theory can only depend on three coordinates $(x_1,x_2,y)$. All the warp factors in the metric and the tensor-structure of the four-form flux respecting the isometries and SUSY preservation can be written---as shown by LLM \cite{Lin:2004nb}, in terms of a function $D(x_1,x_2,y)$ satisfying a Toda equation. Further imposing invariance under $x_1$ translations, we can reduce to Type IIA and after an implicit change of variables $(x_2,y)\to(\sigma,\eta)$, we are left with a Laplace problem in three dimensions with an $x_1$-axial symmetry for a potential function $V(\sigma,\eta)$---see \cite{Lin:2004nb}. 

Reducing in the $x_1$-direction, the Type IIA background and fluxes are completely determined in terms of  $V(\sigma,\eta)$ and its derivatives $\dot{V}=\sigma \partial_\sigma V$, $V'=\partial_\eta V$,  $\ddot V= \sigma\partial_\sigma \dot V$, $\dot V'=\sigma \partial_\sigma V'$ and $V''=\partial_\eta^2 V$. The most general Type IIA configuration with these isometries and preserving 16 SUSYs reads  \cite{Lin:2004nb},
 \be\label{LM}
ds_{10}^2=\left(\frac{\ddot V-2\dot V}{-V''}\right)^{1/2}\left[-4\frac{\ddot V}{\ddot  V-2\dot V}dt^2+\frac{-2V''}{\dot V}(d\sigma^2+d\eta^2)+4d\Omega_5^2+2\frac{V'' \dot V}{\Delta}d\Omega_2^2\right],
\e
with the dilaton, NS-NS and RR potentials given by,
\be\label{LM2}
\begin{split}
e^{4 \phi} =& \frac{4(2 \dot V - \ddot V)^3}{V'' \dot V^2  \Delta^2}, \qquad B_{(2)}= 2 \left(\frac{\dot V \dot V'}{ \Delta} + \eta \right) d \Omega_2,\\
A_{(1)} =& \frac{2\dot V \dot V'}{2 \dot V - \ddot V} dt, \qquad A_{(3)} = - \frac{4 \dot V^2 V''}{ \Delta} dt \wedge d\Omega_2.\\
\end{split}
\e
As we mentioned, the function $V$ solves a Laplace equation (with axial symmetry) 
\be
\ddot V+ \sigma^2 V''=0,\label{laplace}
\e
whose solutions are subject to precise boundary conditions. These were discussed in the paper by Lin and Maldacena \cite{Lin:2005nh}. We can write the potential as a superposition of one part associated with the background electric field and one associated with the particular electrostatic configuration imposed by the boundary conditions, $V(\sigma,\eta)=V_{back}(\sigma,\eta)+ \phi(\sigma,\eta)$.
For the particular case of  the PWMM system, the boundary conditions 
can be summarised as follows:
\begin{itemize}
\item The background electric field must grow at infinity. Imposing that the background asymptotes to the 11d maximally supersymmetric plane wave geometry in eleven dimensions one finds that 
\begin{equation}
\label{back}
V_{back}=V_0(\eta \sigma^2-\frac{2}{3}\eta^3).
\end{equation}

 \item There is 
 an infinite conducting plane at $\eta=0$.
\item There is a set of conducting disks at positions $\eta=\eta_i=\frac{\pi}{2}N_5^{(i)}$, set by the NS5-branes in the background, with radii $R_i$ and charge $Q_i=\frac{\pi^2}{8}N_2^{(i)}$, associated to the D2-brane charges. These NS5 and D2 brane charges are obtained tensoring the $S^2$ and the $S^5$ of the background by different lines in the $(\sigma,\eta)$ plane. Imposing that the electric field on the rim of each disk is finite forces a relation between $R_i$ and $Q_i$. 
\end{itemize}

\begin{figure}[t!]
\begin{varwidth}{\linewidth}
\begin{center}
\begin{tikzpicture}[scale = 0.6]
\draw[thick] (0,0)node[anchor=east]{$0$}[blue] -- (5,0); 
\draw[blue][thick,dotted] (5,0) -- (5.5,0);
\draw[->,thick] (0,-1) -- (0,5); 
\draw[->,thick] (4,-0.6) -- (5,-0.6);

\draw (4.5,-0.47) node[anchor=north]{$\sigma$};

\draw (0,1)node[anchor=east]{\resizebox{.09\hsize}{!}{$\eta_2-\eta_1=\frac{\pi}{2}N_{5}^{(1)}$}}[blue]-- (2,1)[black]node[anchor=west]{\resizebox{.09\hsize}{!} {$Q_1=\frac{\pi^2}{8}N_2^{(1)}$}}
            (0,2) node[anchor=east]{\resizebox{.09\hsize}{!}{$\eta_3-\eta_2=\frac{\pi}{2}N_{5}^{(2)}$}}[blue]-- (3,2)[black]node[anchor=west]{\resizebox{.09\hsize}{!} {$Q_2=\frac{\pi^2}{8}N_2^{(2)}$}}
            (0,3)node[anchor=east]{\resizebox{.09\hsize}{!}{$\eta_4-\eta_3=\frac{\pi}{2}N_{5}^{(3)}$}}[blue] -- (4,3){[black]node[anchor=west]{\resizebox{.09\hsize}{!} {$Q_3=\frac{\pi^2}{8}N_2^{(3)}$}}};

\draw (0,5) node[anchor=east]{$\eta$}
      [blue](2,4) node[anchor=east]{$\vdots$};

 \end{tikzpicture}
\end{center}
\end{varwidth}
\begin{minipage}{0.1\linewidth}
 \begin{align*}
\Longleftrightarrow \quad
 \begin{pmatrix}
  \rotatebox[origin=tl]{-12}
  {$
  \overbrace{\rotatebox{18}{{\small $J_{_{{\tiny N_5^{^{(1)}}}}}$}} \;
  \cdots 
  \rotatebox[origin=c]{18}{{\small $J_{_{{\tiny N_5^{(1)}}}}$}}}^{\rotatebox{18}{{\tiny $N_2^{(1)}$}}}
 \;
  \overbrace{\rotatebox{18}{{\small$ J_{_{N_5^{(2)}}}$}} \;
\;
  \cdots 
  \rotatebox{18}{{\small $J_{_{N_5^{(2)}}}$}}}^{\rotatebox{18}{{\tiny $N_2^{(2)}$}}}
  \; \cdots 
  \overbrace{\rotatebox[origin=c]{18}{{\small $J_{_{{\tiny N_5^{(i)}}}}$}} \;
  \cdots
  \rotatebox{18}{{\small $J_{_{N_5^{(i)}}}$}}}^{\rotatebox{18}{{\tiny $N_2^{(i)}$}}}
  $}
 \end{pmatrix}
\end{align*} 
\end{minipage}
\caption{Electrostatic picture of the PWMM.}
\label{figura1}
\end{figure}
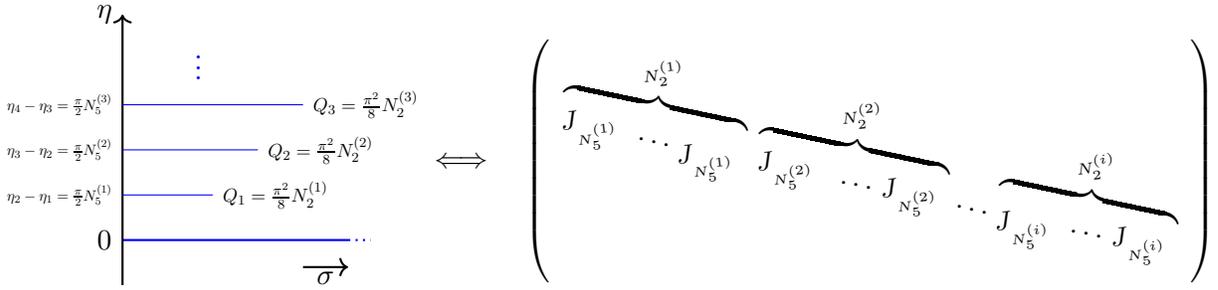

As we pointed out before, the PWMM describes D0-branes plus some relevant deformations. Therefore, the corresponding solutions must asymptote to the metric of D0-branes. This can be seen studying the leading large distance asymptotic form of the solution, which is obtained with a potential  \cite{Lin:2005nh},
\be\label{potpwmm}
V_{_{PWMM}}=V_{back}+P\frac{\eta}{(\eta^2+\sigma^2)^{3/2}},
\e
which includes a 'dipole' sub-leading correction, produced by the disks and their images with respect to the conducting plane at $\eta=0$.
Plugging eq.(\ref{potpwmm}) into the metric given by eq.(\ref{LM}), the asymptotics at $(\eta,\sigma) \rightarrow\infty$ becomes 
\begin{equation}
ds^2\sim -h^{-1/2}dt^2+2^{4}h^{1/2}\left(d\eta^2+\eta^2d\Omega_2^2+d\sigma^2+\sigma^2d\Omega_5^2\right),
\end{equation}
which, changing coordinates  $r^2=16(\eta^2+\sigma^2)$ and considering the limit of very large distances, $r\rightarrow \infty$, becomes the standard D0-brane solution
\begin{equation}
\label{d0}
ds^2\sim -h^{-1/2}dt^2+2^{4}h^{1/2}(dr^2+r^2d\Omega_8^2), \qquad h\sim\frac{2^8 15 P}{r^7}.
\end{equation}
The number of D0-branes is thus proportional to $P$, the dipole moment
produced by the disks, given by $P=2\sum_i \eta_i Q_i \sim \sum_i (\sum_{j<i} N_5^{(i)})N_2^{(i)}$ for a distribution of $i$ disks. Comparing this with eq.(\ref{breaking}) we see that  $N_5^{(i)}$ is related to the dimension of the irreducible representations and $N_2^{(i)}$ with the number of copies defining the BMN vacuum, as illustrated in Figure \ref{figura1}.

In summary \cite{Lin:2005nh}, in order to find the gravity duals of the different vacua of the PWMM, we need to search for a solution to the Laplace equation in the $(\sigma,\eta)$-space with a set of finite size $R_i$ conducting plates positioned at distances $\eta_i=\frac{\pi}{2}N_5^{(i)}$ and  one infinite conducting disk at $\eta=0$. The charges of the finite plates satisfy $Q_i =\frac{\pi^2}{8} N_2^{(i)}$. These charges are carefully chosen  to produce a smooth geometry. Such solution of the Laplace problem---when replaced in eqs.(\ref{LM})-(\ref{LM2}), results in a Type IIA background that is conjectured to be dual \cite{Lin:2005nh} to the vacuum of the PWMM  described by $N_2^{(i)}$ copies of dimension $N_5^{(i)}=(2 j_i+1)$ irreducible representations. Figure \ref{figura1} summarises this relation. 

Notice that in principle, we could have tunneling between different vacua\footnote{Tunneling between different BMN vacua has been studied in \cite{Lin:2006tr}.}, but in the 't Hooft limit with large coupling this is suppressed (as $e^{-\frac{N}{\lambda_t}}$, as it is mediated by instantons) and it is justified to discuss solutions referring to a particular vacuum, as above.

We will now sketch different methods developed in the literature to solve the electrostatic problem described above.

\subsection{Solutions to the Laplace Problem}\label{Laplace}
Different papers dealt with the challenging problem of solving the Laplace equation in the $(\sigma,\eta)$ plane with the boundary conditions imposed by a configuration of one infinite and many finite size conducting disks. As we explained above, any solution of such problem captures the strong coupling dynamics of the PWMM in a particular vacuum.

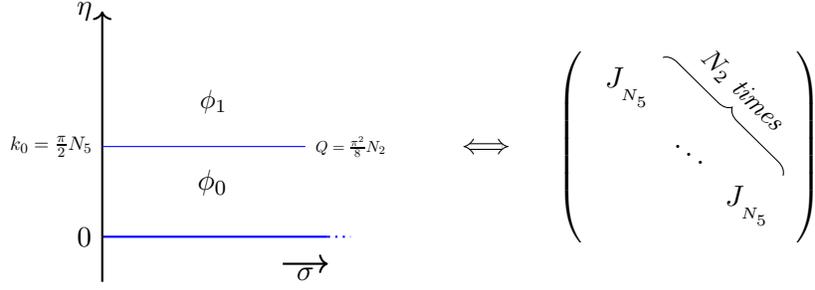
\begin{figure}[t!]
 \[ 
   \begin{tikzpicture}[decoration={brace,amplitude=5pt},baseline=(current bounding box.west),scale = 0.6] 
   \draw[thick] (-13,-2)node[anchor=east]{$0$}[blue] -- (-8,-2); 
\draw[blue][thick,dotted] (-8,-2) -- (-7.5,-2);
\draw[->,thick] (-13,-3) -- (-13,3); 
\draw[->,thick] (-9,-2.6) -- (-8,-2.6);

\draw (-8.5,-2.47) node[anchor=north]{$\sigma$};

\draw (-13,0)node[anchor=east]{\resizebox{.07\hsize}{!}{$k_0=\frac{\pi}{2}N_{5}$}}[blue] -- (-8.5,0){[black]node[anchor=west]{\resizebox{.06\hsize}{!} {$Q=\frac{\pi^2}{8}N_2$}}};

\draw (-13,3) node[anchor=east] {$\tiny{\eta}$}
      (-10,1) node[anchor=east]{{\small $\phi_1$}}
      (-10,-0.8) node[anchor=east]{$\phi_0$}; 
    \matrix (magic) [matrix of math nodes,left delimiter=(,right delimiter=)] {
      J_{_{_{N_{_{5}}}}} \\
       & \ddots \\
      & & J_{_{_{N_{_{5}}}}} \\
   };
   [decoration={brace,amplitude=5pt},baseline=(current bounding box.west)]
   \draw[implies-implies,double equal sign distance] (-5,0) -- (-4,0);
     \draw[decorate] (magic-1-1.north east) -- (magic-3-3.north east) node[above=4pt,midway,sloped]{{\small $N_2$ {\it times } }};
   \end{tikzpicture}
\]
\caption{Electrostatic problem associated to $N_2$ copies of $N_5$-dimensional representations.}
\label{discos2}
\end{figure}

Considering the situation depicted in  Figure \ref{discos2}, in which the vacuum is described by $N_2$ copies of representations of dimension $N_5$, the dual electrostatic distribution of conductors consists on a plate placed at $\eta=\eta_0=\frac{\pi}{2}N_5$, that extends in the $\sigma$-direction up to a distance $\sigma=R$ (the potential on such plate is taken to be $V=\Delta$), plus an infinitely long conducting plate at $\eta=0$, where by convention, we choose to set $V=0$. Writing the electrostatic potential function as
$V=V_{back}+\phi(\sigma,\eta)$, with $V_{back}$ given by (\ref{back}) and $\phi(\sigma,\eta)$ satisfying the Laplace equation, it must satisfy continuity on the disk 
$V(0\leq\sigma\leq R,\eta=k_0-\epsilon)= V(0\leq\sigma\leq R,\eta=k_0+\epsilon)=\Delta$, and continuity of its derivative outside the disk, 
$\partial_\eta V(R\leq\sigma,\eta=k_0-\epsilon)=\partial_\eta V(R\leq\sigma,\eta=k_0+\epsilon)$.

A solution for the function $\phi(\sigma,\eta)$ satisfying these conditions can be obtained by separating variables.  This leads to solutions in terms of exponentials and a Bessel function, basically a Fourier-Bessel transform for each of the two regions depicted in Figure \ref{discos2},  $\phi_l\sim \int_0^\infty
(A_l(u) e^{- u \eta}+ B_l(u)e^{u\eta})  J_0(u\sigma)$, for $l=1,2$. Following \cite{Ling:2006up} and \cite{vanAnders:2007ky}, for an efficient way of imposing the conditions above and after lengthy algebra, a pair of coupled integral equations and  relations among the functions $A_l(u),B_l(u)$ can be obtained.
 
The resolution of these integral equations  is quite challenging. For the case at hand, they were solved  implicitly in \cite{Ling:2006up} and  \cite{vanAnders:2007ky} in terms of nested integrals.
The problem was generalised to the case of many disks and solved---again in an implicit fashion and in terms of nested integrals-- by the authors of \cite{Asano:2014vba}. 

The main message is that finding analytic solutions of the  {\it precise} Laplace problem dictated by the Lin-Maldacena formalism, is quite involved. In the following, we will study an approximated version of the Lin-Maldacena problem for which it is easier to find solutions, and  we will discuss the physical implications of such approximation.

This approximation was originally devised by the authors of the paper \cite{Bak:2005ef} (written before the  work of Lin and Maldacena!). Whilst they analise exact solutions to the Laplace equation (\ref{laplace}), the 'approximation' takes place in the boundary conditions. 
Instead of  a discrete distribution of plates with finely-tuned charges $Q_i$, placed at  precise positions  $\eta_i$ and with radii  $R_i$,  the solutions in \cite{Bak:2005ef} describe the situation in which on the $\eta$-axis, we have a continuous distribution $Q(\eta)$ of very short charged conductors. As explained in \cite{Bak:2005ef}, for the $S^2$ and the $S^5$ to shrink smoothly it is necessary that  $Q(\eta)$ is odd and positive, but this is not sufficient to avoid  a null singularity, that appears in the geometry at the point where the radial component of the electric field vanishes, and is rather similar to that of extremal Dp-branes. 
Subsequently, the authors of \cite{Shieh:2007xn,Donos:2010va}, confirmed this and gave an interpretation in terms of a coarse-grained (or Thermodynamical) description around a 'typical state' (or more probable arrangement of conducting plates). Interestingly, the presence of the singularity in these backgrounds is a consequence of this
'average' or coarse-graining of the Lin-Maldacena distribution of conducting plates, as it happens 
for microstates of black holes, discussed in \cite{Balasubramanian:2005mg}.

Following  \cite{Bak:2005ef}, the electrostatic problem is then reduced to finding a solution of the Laplace equation, subject to the condition that an infinite conducting plane lays at $\eta=0$ and a continuous charge distribution $Q(\eta)$ is located on the $\eta$-axis. Using for instance the method of images one gets that the solution for such electrostatic problem is ---see \cite{Bak:2005ef,Shieh:2007xn,Donos:2010va},
\begin{equation}
V(\sigma,\eta)= V_0 (\eta\sigma^2-\frac{2}{3}\eta^3) +\int_0^\infty dz Q(z)\big[ \frac{1}{\sqrt{\sigma^2+(z-\eta)^2}} -\frac{1}{\sqrt{\sigma^2+(z+\eta)^2}}  \big].
\label{potentialcont}
\end{equation}
Given that in the solution  $\int_0^\infty Q(\eta) d\eta$ counts the number of D2-branes and $\int_0^\infty \eta Q(\eta) d\eta$ the number of D0-branes,
the proposal of \cite{Bak:2005ef} is to make a correspondence between this function and the partitions of $N$---that we called $n_k$ in eq.(\ref{breaking}), such that
\begin{eqnarray}
\label{QQ}
& & \int_0^\infty Q(\eta) d\eta=\sum_k^T n_k,\;\;\;\; \int_0^\infty \eta Q(\eta) d\eta= \sum_k^T k n_k.
\end{eqnarray}
In other words, the function $Q(\eta)$ is giving a 'density' of $SU(2)$ representations of dimension $\eta$. Somewhat  as it happens for the case of four dimensional ${\cal N}=2$ SCFTs,  the $\eta$-direction in the dual gravity problem becomes the 'theory space' direction, see for example \cite{Lozano:2016kum}.
This interpretation was confirmed by other arguments in  \cite{Shieh:2007xn,Donos:2010va}.

Let us close this section with a brief summary. The method in  \cite{Bak:2005ef} allows to find a solution to the Laplace equation (\ref{laplace}). This solution does not satisfy the boundary conditions that a Lin-Maldacena problem demands, hence this leads to singularities in the generated background. Indeed, instead of a discrete set of plates, with precisely defined positions and charge-to-radius relations, a continuous distribution of charge is considered. From the perspective of the  field theory, one loses the precise dimensions of each entry in the vacuum state matrices, to just keep a 'density' of $SU(2)$ representations represented by $Q(\eta)$.
This is the method we will use to understand and make sense of the backgrounds generated by non-Abelian T-duality. We now move on to discussing this new background.

\section{A new non-Abelian T-dual to $AdS_5\times S^5$ }\label{sectionNATD}
In this section, we study the background generated by applying  a non-Abelian T-duality transformation to the $AdS$ subspace of $AdS_5\times S^5$. This will be a {\it new} solution of the Type IIA equations of motion that can be understood in the context of dual backgrounds to PWMM vacua. This connection will provide us with a precise way of understanding and curing problematic features of our new solution.

Let us start by considering the background  $AdS_5\times S^5$, with the $AdS_5$-subspace written in global coordinates, with time and radial-coordinates $(t,r)$. Let us use the left invariant forms of $SU(2)$ to write the three-sphere as $ d\Omega_3^2=\sum_{i=1}^3 \frac{\omega_i^2}{4}$, where,
\begin{equation}
\omega_1=\cos\psi d\varphi+\sin\theta \sin\psi d\varphi,\;\; \omega_2=-\sin\psi d\varphi+\sin\theta \cos\psi d\varphi,\;\;\omega_3=d\psi+\cos\theta d\varphi.\nonumber
\end{equation}
In terms of these, the background reads, 
\be
\label{ads5}
\begin{split}
ds^2=&L^2\left(-\cosh^2 r \,dt^2 + dr^2+\frac{\sinh^2 r}{4}(\omega_1^2+\omega_2^2+\omega_3^2)+d\Omega_5^2\right),\\
&F_5=\frac{4}{L}\Big(Vol(AdS_5)-d\Omega_5\Big).
\end{split}
\e
The parameter $L$ is the radius of $AdS_5$ (also the size of the five sphere) and obeys the condition $\frac{L^4}{\alpha'^2}=4\pi g_s N_{D3}$, that quantises the size of the space.

Let us perform a non-Abelian T-duality transformation using one $SU(2)$ inside the $SO(4)$ isometry of the previous three sphere. The resulting configuration is a new Type IIA background that reads,
\be
\begin{split}\label{natdads}
\frac{ds^2}{L^2}=&-\cosh^2 r \, dt^2+dr^2+\frac{4\alpha'^2}{L^4 \sinh^2 r}d\rho^2+\frac{4\alpha'^2\rho^2 \sinh^2 r}{16\alpha'^2\rho^2+L^4\sinh^4 r}d\Omega_{2}^2(\chi,\xi)+d\Omega_5^2,\\
B_2=&\frac{16\alpha'^3\rho^3}{16\alpha'^2\rho^2+L^4\sinh^4 r}d\Omega_{2}(\chi,\xi),\quad A_1=-\frac{L^4}{8\alpha'^{3/2}}\sinh^4 r dt,\quad F_2=dA_1,\\
e^{-2\phi}=&\frac{L^2\sinh^2r}{64\alpha'^3}(16\alpha'^2\rho^2+L^4\sinh^4 r),\quad F_4=dA_3+A_1\wedge H_3=B_2\wedge F_2.
\end{split}
\e
This solution is singular at the point $r=0$, due to the presence of NS five branes. Indeed, close to $r\sim 0$ the metric and dilaton become 
\begin{equation}
ds^2\sim -dt^2+d\Omega_5^2+\frac{1}{4\nu}(\frac{16 \alpha'^2}{L^4}d\rho^2+d\nu^2+\nu^2 d\Omega_2^2), \,\,\,
e^\phi\sim \frac{2\sqrt{\alpha'}}{L\rho \sqrt{\nu}},\label{qqq}
\end{equation}
with $\nu=r^2$, which is the metric of smeared NS five branes, see \cite{Lozano:2016wrs}.
In Section \ref{QFT}, we will make sense of this singular behaviour using the field theory information reviewed in Section \ref{pwmm}. 

Note that the previous singularity could also be avoided in the presence of a black hole event horizon. Indeed, we could have started with the black hole in global $AdS_5$. In this case we have the same RR flux plus a metric  that reads,
\begin{eqnarray}
& & ds^2=L^2\left(-h\cosh^2 r \,dt^2 + h^{-1}dr^2+\frac{\sinh^2 r}{4}(\omega_1^2+\omega_2^2+\omega_3^2)+d\Omega_5^2\right),\label{bhads}\\
& & h=1+\frac{c_1}{\sinh^2 r}+\frac{c_2}{\cosh^2 r},\nonumber
\end{eqnarray}
where the function
$h(r)$
is the blackening factor and $c_{1},c_{2}$  are constants. The event horizon is
localised at the point where $h(r_h)=0$. 
One can see that the blackening factors $h(r)$ and $h^{-1}(r)$ appear in the $SU(2)$ NATD solution associated to this background in front of the $g_{tt}$ and $g_{rr}$ metric components, while the RR and NS-NS fields remain the same\footnote{The non-Abelian T-dual of the global-AdS black hole was considered in unpublished material developed with  Anibal Sierra and  Dimitrios Giataganas.}. Further study of this solution could be useful in order to understand finite temperature effects. In this paper we will focus however on the zero temperature background described by eq. (\ref{natdads}) without horizons.

Let us then consider the  solution in eq.\eqref{natdads}. This is a half-BPS
background (it preserves 16 supercharges) of Type IIA supergravity with $\mathbb{R}_t\times SO(3)\times SO(6)$ isometry. Therefore it should belong to the general class of solutions described by 
eqs.(\ref{LM})-(\ref{laplace}). Indeed, one can check that with the change of coordinates,
\be\label{changeeta}
\sigma=\cosh r,\qquad \eta=\frac{2\alpha'}{L^2}\rho,
\e
we can derive the solution in  eq.\eqref{natdads} from a potential\footnote{A suitable rescaling of parameters is necessary to match the results of \cite{Lin:2005nh}.},
\be\label{vv}
\begin{split}
V_{_{\textrm{NATD}}}=&\eta\left(-\log\sigma+\frac{\sigma^2}{2}\right)-\frac{\eta^3}{3}=\frac{1}{2} (V_{_{\textrm{back}}}-2\eta \log \sigma),
\end{split}
\e
where $V_{_{\textrm{back}}}$ is the background potential associated to the PWMM. Indeed,  
in these coordinates, the background metric reads,
\begin{equation}
ds^2=-4\sigma^2dt^2+\frac{4}{\sigma^2-1}(d\sigma^2+d\eta^2)+\frac{4\eta^2(\sigma^2-1)}{4\eta^2+(\sigma^2-1)^2}d\Omega_2^2+4d\Omega_5^2,\label{xxy}
\end{equation}
which can be checked to be that of eq.(\ref{LM}) for the potential $V_{_{\textrm{NATD}}}$ above. 
Similar expressions can be calculated for the fluxes and dilaton in ($\sigma,\eta$)-coordinates, using eq.(\ref{natdads}), the change in eq.(\ref{changeeta}) and comparing them to those in eq.(\ref{LM2}) for $V_{\textrm{NATD}}$.

In light of what we discussed in  Section \ref{pwmm}, the singularity of the non-Abelian T-dual solution 
at $\sigma=1$ should then be a consequence of the fact that our background does not satisfy the boundary conditions 
itemised  below eq.(\ref{laplace}). We will come back to this issue in the following subsections.
Let us now proceed with a detailed study of our new Type IIA solution.

  \subsection{Asymptotics} \label{asymptotics}
As we have pointed out, the PWMM describes D0 branes plus some relevant deformations. Therefore, the corresponding solutions must asymptote to the metric of D0 branes. Let us analyse whether this is the case for the non-Abelian T-dual solution.


The logarithmic term in the potential associated to the non-Abelian T-dual solution,  given by eq. (\ref{vv}), does not vanish at infinity, unlike the sub-leading dipolar deformation of eq.(\ref{potpwmm}). This suggests that the non-Abelian T-dual solution will not have D0-brane asymptotics. This is indeed the case. Consider 
 the limit $\sigma\rightarrow \infty,~\eta=\textrm{fixed}$, in the non-Abelian T-dual background metric. The metric in eq.(\ref{xxy}) becomes,
\be\label{infnatd}
ds^2\sim-4\sigma^2 dt^2+\frac{4}{\sigma^2}(d\eta^2+d\sigma^2)+4\frac{\eta^2}{\sigma^2}d\Omega_2^2+4d\Omega_5^2.
\e
In order to understand what brane configuration gives this asymptotics, consider D0 branes defined by the metric in eq. \eqref{d0}.
If we smear the D0 branes on the $(\eta,\Omega_2)$ directions, which implies a continuous superposition of these D0 branes \cite{HoyosBadajoz:2010td, Faedo:2013aoa}, we find
\be
\begin{split}
h_{\textrm{s}}(\sigma)=&\int_{0}^{\infty}\eta^2d\eta\int d\Omega_2h_{l}(\sigma,\eta)
\sim 4\pi\int_{0}^{\infty}\frac{\eta^2 d\eta}{(\eta^2+\sigma^2)^{7/2}}\sim\frac{8\pi}{15\sigma^4},
\end{split}
\e
where $h_{s,l}$ denote the warp factor of the smeared and localised D0 brane solutions respectively. 
Then, the smeared D0 brane metric behaves like, 
\be
ds^2_{s}\sim-\sigma^2 dt^2+\frac{1}{\sigma^2}\left(d\eta^2+\eta^2d\Omega_2^2+d\sigma^2+\sigma^2d\Omega_5^2\right), \nonumber
\e
which is precisely the asymptotic form of the NATD solution in eq. \eqref{infnatd}. Therefore, the large-$\sigma$ asymptotics of the NATD solution is that of smeared D0 branes on $\mathbb{R}^3$.
Had we considered the situation  in which the D0 branes had been smeared on the finite range 
$\eta\in [0,\Delta^{7/3}]$, we would have found a smeared warp factor,
$h_{\textrm{s}}(\sigma)\sim4\pi\int_{0}^{\Delta^{7/3}}\frac{\eta^2 d\eta}{(\eta^2+\sigma^2)^{7/2}}\sim\left(\frac{\Delta}{\sigma}\right)^7$,
which has the same large distance asymptotics as the D0 brane metric in eq.\eqref{d0}. 

The conclusion is that the non-Abelian T-dual solution of eq. \eqref{infnatd} can be interpreted  as an infinite superposition of D0 branes. Its different asymptotic behaviour can then be associated to the presence of  an irrelevant operator in the matrix model, whose effect is to change the dimensionality of the field theory. Given that an irrelevant operator is never welcomed in a field theory, we will use the ideas summarised in Section \ref{pwmm} to 'cure' this  problem.

Similarly to the analysis above, we can consider the limit in which $\sigma$ is kept fixed, but the coordinate $\eta\to\infty$. We can see that in this case the NATD background  asymptotes to the Abelian T-dual (in the $U(1)_\psi$-direction) of the solution in eq.(\ref{ads5}). This connection between non-Abelian T-duality and its Abelian counterpart was observed in \cite{Macpherson:2015tka} and later refined in \cite{Lozano:2016kum}. Here  again, the metric represents a dual to a quantum field theory in  dimension higher than $(0+1)$. 
 %

\subsection{Realisation as a smeared Lin solution}
\label{lin}
In the previous subsection, we have shown that our new Type IIA background obtained by 
applying non-Abelian T-duality is a singular solution, and that it can be understood as describing D0 branes that have been smeared along $\mathbb{R}^3$.

In this section we will make contact with a nice work by Hai Lin \cite{Lin:2004kw}, where,
applying the method used by Polchinski and Strassler \cite{Polchinski:2000uf}  to construct the Type IIB duals to the vacua of the 
$\mathcal{N}=1^*$ field theory, he considered backgrounds of Type IIA supergravity obtained by perturbing the near horizon geometry of $N$ D0-branes by transverse RR 6-form and NS-NS 3-form fluxes. These fluxes are the Type IIA duals to the deformation of the BFSS matrix model by a mass and a dielectric term, thus giving rise to the Type IIA realisation of the BMN matrix model \cite{Berenstein:2002jq}. Lin showed that in the presence of these fluxes the D0-branes polarise to concentric D2-branes with radii proportional to their D0-charge. When lifted to eleven dimensions, these solutions describe  giant gravitons that expand into M2-branes on the eleven-dimensional pp-wave background, thus making contact with the ideas proposed by  \cite{Bak:2005ef}, subsequently refined by
the authors of \cite{Shieh:2007xn,Donos:2010va}.
We will show that our NATD solution solves the equations proposed by
  \cite{Lin:2004kw}, provided that the branes are smeared. This, together with the material discussed in Section
  \ref{asymptotics}, reinforces the interpretation of the NATD solution in terms of smeared D0 branes.

Let us start by briefly  summarising the main results of the work \cite{Lin:2004kw}. 
The Type IIA solutions constructed in \cite{Lin:2004kw} are perturbations in the large-$r$ limit around the near horizon geometry of $N$ D0-branes, by magnetic RR 6-form and NS-NS 3-form fluxes.  The background metric, dilaton and $F_2$ flux characterising D0-branes, 
\begin{eqnarray}
ds^2=-h^{-1/2}dt^2+h^{1/2} (dr^2+r^2 d\Omega_8^2)\,;\;\; e^\Phi=h^{3/4} ,\;\;A_1=h^{-1} dt,\;\; h=\frac{R^7}{r^7}, \;\; R^7=60\pi^3 N\, ,\nonumber
\end{eqnarray}
are perturbed up-to-second order in the fluctuations of the fields $H_3, F_6$. One just needs to solve the equations of motion for the fluxes, which take the simple form \cite{Lin:2004kw},
\begin{eqnarray}
\label{linearfluxes}
&&dH_3=0\nonumber\\
&&dF_6=0\nonumber\\
&&d[h^{-1} (H_3-*_9 F_6)]=0\nonumber\\
&&d[h^{-1}(*_9 H_3-F_6)]=0
\end{eqnarray}
where in these expressions $*_9$ is the Hodge dual in the transverse nine-dimensional space with respect to a flat 9d metric. 

Let us now come back to the $\sigma\rightarrow \infty$ asymptotic behaviour of the non-Abelian T-dual solution. As we showed below eq.(\ref{infnatd}), the behaviour of the metric is that of smeared D0-branes on $\mathbb{R}^3$. The  same smearing happens for the dilaton and $A_1 $ RR field (they are also written in terms of the smeared warp factor $h_s$) that read,
\begin{eqnarray}
&&e^\Phi\sim \frac{8 \alpha'^{3/2}}{L^3}\sigma^{-3}\, ,\qquad A_1\sim -\frac{L^4}{8\alpha'^{3/2}}\sigma^4 dt.\nonumber
\end{eqnarray}
Thus, as a result of the smearing, the fields $H_3$,  $F_6$  and their $*_9$-duals are given to leading order in $\sigma$ as,
\begin{eqnarray}
&&H_3\sim \frac{6L^2\eta^2}{\sigma^4}(d\eta-\frac43 \frac{\eta}{\sigma} d\sigma)\wedge d\Omega_2,\;\;\;F_6=-\frac{L^8}{8\alpha'^{3/2}} \eta d\eta\wedge d\Omega_5,\nonumber\\
&&*_9 F_6\sim\frac{L^5}{4\alpha'^{3/2}} \frac{\eta^3}{\sigma^5}d\sigma\wedge d\Omega_2,\;\;\;*_9 H_3\sim-L^5 (4\eta d\eta+3\sigma d\sigma)\wedge d\Omega_5.\nonumber
\end{eqnarray}
Using these fluxes, one can show that the first two lines of  eq.(\ref{linearfluxes}) are straightforwardly satisfied. Considering that the warp factor is a function of $\sigma$ only, $h(\sigma)$, one can show that the fourth equation in (\ref{linearfluxes}) is satisfied, while the third line of eq.(\ref{linearfluxes})  imposes that $h^{-1}\sim \sigma^4$. This is, precisely,  the same result as the smearing of the D0 branes on $\mathbb{R}^3$.


In summary, we have shown that the
non-Abelian T-dual  configuration provides a full solution (beyond the perturbative expansion devised by Lin) for a background of smeared  D0 branes deformed by asymptotically sub-leading  $F_6$ and $H_3$ fluxes. 
There remain the unpleasant features of the smearing along a non-compact space (which  obscures the
field theoretical interpretation of the non-Abelian T-dual solution), and 
the fact that the background is singular at $\sigma=1$ (which can be physically interpreted following  \cite{Bak:2005ef}, \cite{Shieh:2007xn} and \cite{Donos:2010va}).
We will discuss this in Section \ref{QFT}. Before that and to get more clues for the field theoretical understanding of the non-Abelian T-dual solution, we will calculate its conserved charges.
 \subsection{Quantised charges} \label{charges}
 The study of the quantised Page charges was instrumental in
 developing the field theory dual of the only backgrounds obtained via non-Abelian T-duality to which a concrete CFT dual has been proposed  
  \cite{Lozano:2016kum,Lozano:2016wrs}. In these examples, the quiver associated with the field theory was encoded in a brane set-up compatible with the Page charges of the solution.
 In this section, we study the conserved charges of our Type IIA background. 

 The presence of $F_2$ and $F_4$ fluxes suggests that the solution should be associated with a  combination of  D0 and  D2-branes. The Hodge duals of the fluxes  in eq.(\ref{natdads}) are given by, 
  \begin{equation}
 \label{F6F8}
 F_6=-4\sqrt{\alpha'}L^4 \rho d\rho \wedge d\Omega_5\, , \quad
 F_8=\frac{4L^8\alpha'^{3/2} \rho^2 \sinh^4{r}}{16\alpha'^2 \rho^2+L^4 \sinh^4{r}} d\rho\wedge d\Omega_2\wedge d\Omega_5.
 \end{equation}
 The relevant ``Page flux'' $\hat{F}_8=F_8-B_2\wedge F_6$  reads in turn,
 \begin{equation}
 {\hat F}_8=4L^4 \alpha'^{3/2}\rho^2 d\rho \wedge d\Omega_2\wedge d\Omega_5.
 \label{pagef8}
 \end{equation}
 
 
In analogy with other cases studied in the literature---see for example  \cite{Lozano:2016kum, Lozano:2013oma,Lozano:2016wrs,Macpherson:2014eza,Lozano:2014ata}-- the  $B_2$-field should be defined up-to large gauge transformations, associated to the non-trivial 2-cycle that sits at the singularity at $r=0$, where the cone spanned by $(r,\Omega_2)$ becomes singular. Indeed,  $B_2$ should be defined as,
 \begin{equation}\label{B2field}
 B_2=\Bigl(\frac{16\alpha'^3\rho^3}{16\alpha'^2\rho^2+L^4\sinh^4 r}-\alpha'k\pi\Bigr)d\Omega_{2}(\chi,\xi),
 \end{equation}
 and this naturally divides the $\rho$-coordinate in intervals of size $\pi$, with a large gauge transformation as that in eq.(\ref{B2field}), performed every time the interval $\rho\in [k\pi,(k+1)\pi]$ is crossed. This affects $\hat{F}_8$  in eq.(\ref{pagef8}) as,
 \begin{equation}
 \hat{F}_8\rightarrow  \hat{F}_8=4\alpha'^{3/2}L^4\rho (\rho -k\pi) d\rho\wedge d\Omega_2\wedge d\Omega_5.
 \end{equation}
 
Using that $2\kappa_{10}^2 T_{Dp}= (2\pi)^{7-p} g_s \alpha'^{(7-p)/2}$ and setting $g_s=1$,
the D2 and D0 charges associated to $F_6$ and $\hat{F}_8$  read, for $\rho\in [k\pi,(k+1)\pi]$,  
\begin{eqnarray}
 \label{D2charge}
&& N_2=\frac{1}{(2\pi)^5 \alpha'^{5/2}}\int_{\rho,\Omega_5} F_6=-\frac{4L^4\Omega_5}{(2\pi)^5 \alpha'^2}\int_{k\pi}^{(k+1)\pi}\rho d\rho=
 -\frac{L^4}{16\alpha'^2}(2k+1) \\
 &&N_0=\frac{1}{(2\pi)^7 \alpha'^{7/2}}\int_{\rho,\Omega_5,\Omega_2} {\hat F}_8=\frac{4L^4 \Omega_5 \Omega_2}{(2 \pi)^7 \alpha'^2}\int_{k\pi}^{(k+1)\pi}\rho(\rho-k\pi)d\rho=\frac{L^4}{8\alpha'^2}(\frac{k}{2}+\frac13)
 \label{D0charge}
 \end{eqnarray}
We have also NS five brane charge, 
 \begin{equation}\label{H3}
 N_5=\frac{1}{(2\pi)^2\alpha'}\int_{\rho,S^2}H_3=\frac{1}{\pi}\int_{0}^{k\pi} d\rho=k.
 \end{equation}
 The NS five branes are located at $\rho=\pi, 2\pi,\dots k\pi$, and $r=0$ (note that we integrate over the $S^2$ at $r=0$).
 There is thus no brane set-up in which color branes are stretched between pairs of NS five branes. This represents an essential difference with respect to previous backgrounds constructed through non-Abelian T-duality, such as the examples in  \cite{Lozano:2016kum,Lozano:2016wrs}.
 
 Our discussion in Section \ref{lin} suggests that the correct way to interpret the D0 and D2 brane charges is instead in terms of D0-branes expanded onto spherical D2-branes. Indeed, the second term in the integral in equation (\ref{D0charge}) originates from 
 \begin{equation}
 \label{D0dipolar}
N_0^{(d)}= \frac{k\pi}{(2\pi)^7}\int_{\rho,S^5}F_6\int_{S^2} d\Omega_2,
 \end{equation}
 and is thus equal to, 
 \begin{equation}
 \label{D0dipolar2}
 N_0^{(d)}=-\frac{L^4}{16\alpha'^2}k(2k+1)=kN_2\, ,
 \end{equation}
with $N_2$ the D2-brane charge given by eq.(\ref{D2charge}). Thus, this contribution to the D0-brane charge comes from the D0-branes that are dissolved in the D2-branes. Conversely, we can think of the D2-branes as originating from D0-branes expanded onto fuzzy 2-spheres through Myers' dielectric effect \cite{Myers:1999ps}. Indeed, 
D2-branes in the non-Abelian T-dual background only carry D0-brane dipole charge, induced by the large gauge transformations of the $B_2$ field (see the next section). We have used the notation $N_0^{(d)}$ in equations (\ref{D0dipolar}), (\ref{D0dipolar2}) to explicitly  account for this contribution to the total D0-brane charge in the configuration. The remaining charge at each interval,  
\begin{equation}
\label{D0monopole}
N_0^{(m)}=\frac{4L^4 \Omega_5\Omega_2}{(2\pi)^7\alpha'^2}\int_{k\pi}^{(k+1)\pi} \rho^2 d\rho=\frac{L^4}{24\alpha'^2}(3k^2+3k+1)
\end{equation}
is just standard (monopole) D0-brane charge.


\begin{figure}[t!]
\centering
\begin{tikzpicture}
\coordinate (O) at (0,0);

\draw (O) circle (2.8);
\draw (O) circle (1.25);
\draw (O) circle (0.5);
\draw (0,0.9)node[anchor=north]{\resizebox{.03\hsize}{!}{$1 \,D_0 $}};
\draw (0,1.7)node[anchor=north]{\resizebox{.03\hsize}{!}{$2 \,D_0 $}};
\draw (0,2.7)node[anchor=north]{\resizebox{.008\hsize}{!}{$\vdots$}};
\draw (0,3.3)node[anchor=north]{\resizebox{.03\hsize}{!}{$k \,D_0 $}};
\draw (0.9,0)node[anchor=east]{\resizebox{.01\hsize}{!}{$\pi $}};
\draw (1.8,0)node[anchor=east]{\resizebox{.019\hsize}{!}{$2\pi $}};
\draw (3.3,0)node[anchor=east]{\resizebox{.019\hsize}{!}{$k\pi $}};
\draw (2.5,0)node[anchor=east]{\resizebox{.03\hsize}{!}{$\ldots$}};
\end{tikzpicture}
\caption{Concentric distribution of expanded D0-branes on 2-spheres with radii $\rho=k\pi$. At each radius $k\pi$ there are $N_2=\frac{L^4}{8\alpha'^2} k$ coincident D2-branes.}
\label{concentricD2}
\end{figure}
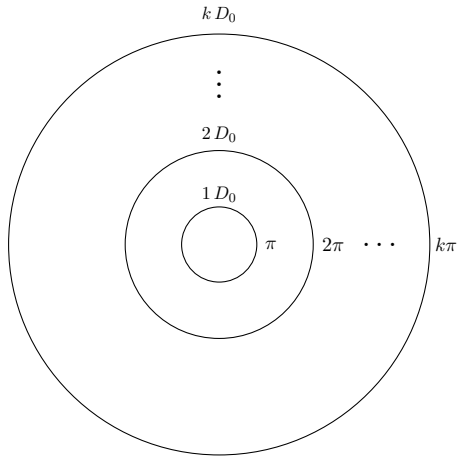

The picture that arises is that of $N_2$ spherical D2-branes with $k$ D0-brane charge dissolved. We will show in the next section that $\rho$ can in fact be interpreted as the radius of smeared D0-branes expanded into the $S^2$ spanned by the coordinates $(\chi,\xi)$ in eq.(\ref{natdads}). The brane set-up is thus a concentric distribution of expanded, smeared, D0-branes on 2-spheres with radii $\rho=k\pi$. This is depicted in Figure \ref{concentricD2}. On top of this there are as well standard point-like D0-branes at $\rho=k\pi$, with charges given by eq.(\ref{D0monopole}), that do not have a direct interpretation in terms of BMN vacua (see below).

Taking a $[0,p\pi]$ interval, the total D0-brane charge distributed among the concentric D2-branes at $\rho=\pi, 2\pi,\dots, p\pi$ is then given by,
\begin{equation}
\label{breaking2}
N=\sum_{k=1}^p k N_2.
\end{equation} 
This can be seen as a partition of the total number of D0-branes of exactly the same form of the partition in eq. (\ref{breaking}), that defines the different BMN vacua. Here $k$ is the dimension of the $SU(2)$ representation, $k=(2j_k+1)$, and $N_2$ its number of copies, $n_k$.  As in \cite{Lin:2005nh}, the dimension of the representations is associated to the number of NS five branes, while their multiplicities are associated to the number of D2-branes of the gravity dual.

Let us discuss in more detail the situation at the $[0,\pi]$ interval. In this interval we have one NS five brane located at $\rho=\pi$ and $N_2=\frac{L^4}{16\alpha^{\prime 2}}$ point-like D2-branes. This configuration should then be associated to the trivial BMN vacuum. It is interesting that this appears to be an explicit realisation of the conjecture in \cite{Maldacena:2002rb} that the trivial vacuum should be associated to a single NS five brane. It would be interesting to understand this connection more precisely.  Accordingly, we have to shift all charges by $\frac{L^4}{16\alpha^{\prime 2}}$ in order to fulfil the condition that  $V=0$ at $\eta=0$ and 
that the D2-brane charge distribution is odd along the $\eta$ axis.
Doing this shift we have $N_2=\frac{L^4}{8\alpha^{\prime 2}} k$ at each $[k\pi, (k+1)\pi]$ interval. Through eq. (\ref{breaking2}), the total D0-brane charge in $[0,p\pi]$ is then
\begin{equation}
\label{numberD0}
N=\frac{L^4}{48\alpha'^2}p(p+1)(2p+1).
\end{equation}

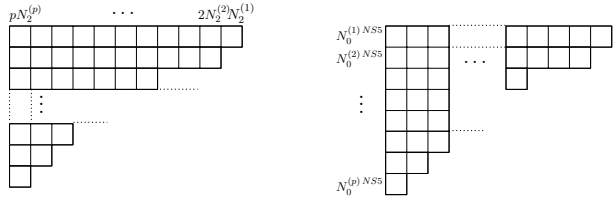
\begin{figure}[t!]
\centering
\begin{tikzpicture}
\Yboxdim{8pt}
\tyng(5 cm, 0cm,3^5)
\tyng(5 cm, -1.4cm,3,2,1)
\tyng(6.6 cm, 0cm,5^1,4^1,1)
\draw [densely dotted](5.84,0.29)--(6.6,0.29);
\draw [densely dotted](5.84,0)--(6.6,0);
\draw [densely dotted](5.84,-1.11)--(6.3,-1.11);
\draw (5.1,0.16)node[anchor=east]{\resizebox{.04\hsize}{!}{$N_{0}^{(1)\, NS5}$}};
\draw (5.1,-0.15)node[anchor=east]{\resizebox{.04\hsize}{!}{$N_{0}^{(2)\, NS5}$}};
\draw (5.1,-1.85)node[anchor=east]{\resizebox{.04\hsize}{!}{$N_{0}^{(p)\, NS5}$}};
\draw (4.85,-0.68)node[anchor=east]{\resizebox{.005\hsize}{!}{$\vdots$}};
\draw (6.48,-0.2)node[anchor=east]{\resizebox{.02\hsize}{!}{$\ldots$}};

\Yboxdim{8pt}
\tyng(0 cm, 0cm,11,10,7)
\draw (0.57,0.45)node[anchor=east]{\resizebox{.028\hsize}{!}{$p N_{2}^{(p)}$}};
\draw (3.08,0.45)node[anchor=east]{\resizebox{.028\hsize}{!}{$2N_{2}^{(2)}$}};
\draw (3.4,0.45)node[anchor=east]{\resizebox{.024\hsize}{!}{$N_{2}^{(1)}$}};
\draw (0.6,-0.68)node[anchor=east]{\resizebox{.005\hsize}{!}{$\vdots$}};
\draw (1.8,0.45)node[anchor=east]{\resizebox{.02\hsize}{!}{$\ldots$}};
\tyng(0 cm, -1.3cm,3,2,1)
\draw [densely dotted](0.85,-1)--(1.3,-1);
\draw [densely dotted](0,-0.56)--(0,-1);
\draw [densely dotted](0.29,-0.56)--(0.29,-1);
\draw [densely dotted](2,-0.56)--(2.5,-0.56);
\end{tikzpicture}
\caption{Young tableaux for the partition (left) and dual partition (right) of N.}
\label{youngtab}
\end{figure}

Finally, following \cite{Maldacena:2002rb}, the same number of D0-branes given by eq.(\ref{numberD0}) should be able to expand at large 't Hooft coupling onto NS five branes wrapped on 5-spheres. The argument presented in  \cite{Maldacena:2002rb} is that we can represent any partition of $N$ by a Young diagram whose column lengths are the elements of the partition. In the D2-brane interpretation such diagram corresponds to a state with one D2-brane for each column, with the number of boxes in the column indicating the D0-brane charge carried by this D2-brane. Alternatively, in the NS five brane interpretation, the rows of the Young diagram correspond to the number of D0-branes carried by each NS five brane. In this dual interpretation the number of NS five branes coincides with the dimension of the largest irreducible representation, and the D0-brane charge carried by the $m$th NS five brane is equal to the number of irreducible representations with a size greater than or equal to $m$. This partition of $N$ is commonly referred as the dual partition. This is depicted in Figure \ref{youngtab}. 
In our case the dual partition to the partition in eq. (\ref{breaking2}) 
describes $p$ NS five branes, with the $m$th of them carrying D0-brane charge given by
\begin{equation}
\label{D0NS5}
N_{0}^{m NS5}=\frac{L^4}{8\alpha'^2}\sum_{k=m}^{p} k=\frac{L^4}{16\alpha'^2} [p(p+1)-m(m-1)]\, ,  \quad m=1,\dots p
\end{equation}
We can check that indeed $N$, as given by eq.(\ref{numberD0}), is reproduced from
\begin{equation}
N=\sum_{m=1}^p N_{0}^{m NS5}.
\label{zzz}
\end{equation}
As we will discuss in Section \ref{dielectric2}, it is possible to construct BPS NS five branes wrapped on the 5-sphere of the non-Abelian T-dual geometry, located at $r=0$ and at given positions in $\rho$, that carry D0-brane charge given by eq.(\ref{D0NS5}).

\subsection{Dielectric D2-branes}
 
 
 As we have discussed, Lin showed in \cite{Lin:2004kw} that in the presence of transverse RR 6-form and NS-NS 3-form fluxes, D0-branes polarise into concentric D2-branes with radii proportional to the D0-charge that they carry. This description is valid in the limit of large D0-charge. 
In this section we show that there are similar concentric D2-branes wrapped on the 2-sphere of the non-Abelian T-dual solution that are interpreted as D0-branes smeared on $r$ and concentric 2-spheres of radii $k\pi$.

Let us consider a probe D2-brane wrapped on the 2-sphere of the non-Abelian T-dual solution,
\begin{equation}
ds^2_{D2}=-L^2 \cosh^2{r}dt^2+\frac{4L^2\alpha'^2\rho^2\sinh^2{r}}{16\alpha'^2\rho^2+L^4\sinh^4{r}}d\Omega_2^2\, ,
\end{equation}
the radius of which depends on both $\rho$ and $r$.
In the presence of large gauge transformations this brane is dielectric and carries D0-brane charge. This can be shown from the couplings to the RR potentials in the WZ action,
\begin{equation}
\label{D2WZ}
S_{D2}^{WZ}=T_2\int [A_3-A_1\wedge B_2],
\end{equation}
where $d(A_3-B_2\wedge A_1)=F_4-B_2\wedge F_2=\alpha'k\pi d\Omega_2\wedge F_2$, and $A_3-B_2\wedge A_1=\alpha'k\pi d\Omega_2\wedge A_1$. Thus, substituting in eq. (\ref{D2WZ}) we see that the brane carries $k$ units of D0-brane charge,
\begin{equation}
\label{D2WZ2}
S_{D2}^{WZ}=k\int_{\mathbb{R}}A_1\, .
\end{equation}
In the particular case of the NATD solution this coupling reads
\begin{equation}
S_{D2}^{WZ}=-k\int_{\mathbb{R}} dt \frac{L^4}{8\alpha'^{3/2}}\sinh^4{r}.
\end{equation}
In order to find the equilibrium position of the brane we compute as well its DBI action:
\begin{eqnarray}
S_{D2}^{DBI}&=&-T_{D2}\int_{\mathbb{R},S^2} d^3\xi e^{-\phi}\sqrt{{\rm det}(g-B_2)}=\nonumber\\
&=&-4\pi T_{D2}\int_{\mathbb{R}}dt \frac{L^2\sinh{r}\cosh{r}}{8\sqrt{\alpha'}}\sqrt{16\alpha'^2\rho^2(\rho-k\pi)^2+L^4\sinh^4{r}k^2\pi^2}.
\end{eqnarray}
A D2 placed in this background has then a potential
\begin{equation}
V(r,\rho)=\frac{L^2}{8\sqrt{\alpha'}\pi}\sinh{r}\cosh{r}\sqrt{16\alpha'^2\rho^2(\rho-k\pi)^2+L^4\sinh^4{r}k^2\pi^2}-\frac{k L^4}{8 \alpha'}\sinh^4{r}
\end{equation}
The equilibrium position of the brane, at which it carries zero energy, is then found
minimising in $\rho$  and $r$. We find that $\rho=k\pi$ and that $r\sim \infty$. The radius of the brane is thus
\begin{equation}
R\sim \frac{2k\pi}{L \sinh{r}}
\end{equation}
which corresponds to a metric
\begin{equation}
ds^2_{D2}\sim -\sigma^2 dt^2+\sigma^{-2} k^2 \pi^2 d\Omega_2^2
\end{equation}
where we have used that $\sinh{r}\sim \cosh{r}\sim \sigma$ for large $r$. The asymptotic behaviour given by this equation corresponds to D0-branes smeared on $r$ and an $S^2$ with radius $k\pi$.  
  
\subsection{Dielectric NS five branes}
\label{dielectric2}

The previous configuration in terms of concentric shells of smeared D0-branes is valid at weak  coupling  \cite{Maldacena:2002rb}. Alternatively, at strong coupling we should find configurations
of NS five branes carrying D0-brane charge. In this section we study these configurations using the effective action describing NS five branes in Type IIA constructed in  \cite{Bandos:2000az}. We refer the reader to this reference for more details on its construction.

Consider a probe NS five brane wrapped on the 5-sphere of the non-Abelian T-dual solution\footnote{Similar configurations of NS five branes wrapped on the $S^5$ were considered in \cite{Lin:2006tr} in the study of tunnelling between different BMN vacua.},
\begin{equation}
ds^2_{NS5}=-L^2\cosh^2{r}dt^2 +L^2d\Omega_5^2.
\end{equation}
This ansatz does not allow configurations of concentric NS five branes, since all branes have the same radius $L$. Instead, we will see that the NS five branes will be located at different positions in the $\rho$ direction according to the D0-charge that they carry.
In order to describe this brane it is enough to use the truncated DBI action in  \cite{Bandos:2000az},
\begin{equation}
S^{DBI}_{NS5}=-T_5\int d^6\xi e^{-2\phi}\sqrt{-{\rm det}(g+e^{2\phi}G_1^2)},
\end{equation}
where $G_1$ is the invariant field strength associated to the RR 1-form potential,
\begin{equation}
G_1=dc_0+A_1,
\end{equation}
through which we will introduce D0-brane charge in the configuration.
Here $c_0$ is the world-volume field that originates in the eleventh scalar of the M5-brane. In close analogy with the more usual BI field,
we can induce D0-brane charge in the configuration taking $\dot{c}_0\neq 0$. 

The relevant part of the WZ action reads in turn:
\begin{equation}
S^{WZ}_{NS5}=T_5\int \Bigl[B_6-A_5\wedge dc_0],
\end{equation}
where $B_6$ is such that
\begin{equation}
d(B_6+A_5\wedge A_1)=e^{-2\phi} *H_3-A_5\wedge F_2 .
\end{equation}
Particularising to the non-Abelian T-dual background,
\begin{equation}
B_6=-\frac{L^8}{64\alpha'^3}\sinh^2{r}(16\alpha'^2\rho^2+L^4\sinh^4{r})
dt\wedge d\Omega_5,
\end{equation}
where we have used  that $A_5=-2\sqrt{\alpha'}L^4 \rho^2 d\Omega_5$ after integrating $F_6$ in eq.(\ref{F6F8}). 

As already mentioned, in order to induce D0-brane charge in the NS five brane we switch on 
$\dot{c}_0\neq 0$. Its conserved conjugate momentum will then give the D0-brane charge carried by the NS five brane.
Indeed, substituting in the action (of a NS five brane in our conventions) we find,
\begin{eqnarray}
S^{DBI}_{NS5}&=&-T_5 \frac{\pi^3 L^6}{64\alpha'^3}\int \sinh{r}\sqrt{16\alpha'^2\rho^2+L^4\sinh^4{r}}.\nonumber\\
 &&.\sqrt{L^4\sinh^2{r}\cosh^2{r}(16\alpha'^2\rho^2+L^4\sinh^4{r})-64\alpha'^3(\dot{c}_0-\frac{L^4}{8\alpha'^{3/2}}\sinh^4{r})^2}
 \end{eqnarray}
 and
 \begin{equation}
 S^{WZ}_{NS5}=T_5 \pi^3\int \Bigl[\frac{L^8}{64\alpha'^3}\sinh^2{r}(16\alpha'^2\rho^2+L^4\sinh^4{r})+2\sqrt{\alpha'}L^4 \rho^2 \dot{c}_0\Bigr].
 \end{equation}
 The corresponding Hamiltonian $H=P \dot{c}_0-L$ reads,
 \begin{eqnarray}
 H=&&T_5 \frac{\pi^3 L^8}{64\alpha'^3}\sinh^2{r}\Bigl[\cosh{r}(16\alpha'^2\rho^2+L^4\sinh^4{r})
 \sqrt{1+\frac{64\alpha'^3(P-2T_5\pi^3\sqrt{\alpha'}L^4\rho^2)^2}{T_5^2\pi^6L^{12} (16\alpha'^2\rho^2+L^4\sinh^4{r})}}- \nonumber\\
&&- (16\alpha'^2\rho^2\cosh^2{r}+L^4\sinh^4{r})\Bigr]+\frac{L^4}{8\alpha'^{3/2}}P\sinh^4{r}\, .
 \end{eqnarray}
 Minimising in $P$ we find that
 \begin{equation}
 \label{Pquant}
 P=2T_5\pi^3\sqrt{\alpha'}L^4\rho^2
 \end{equation}
 and
 \begin{equation}
 H=T_5\frac{\pi^3L^8}{64\alpha'^3}{\sinh^2{r}}\Bigl[(\cosh{r}-1)(16\alpha'^2\rho^2+L^4\sinh^4{r})\Bigr] .
 \end{equation}
 This is in turn minimised for $r=0$, where the NS five brane becomes in fact tensionless.
 
 The output of this analysis is that an NS five brane wrapped on the $S^5$ can be in equilibrium if it is located at the singularity of the background, $r=0$, and at a position in the $\rho$-direction that satisfies that its D0-charge, $P$, is quantised, as imposed by eq.(\ref{Pquant}). This condition shows that NS five branes located at $\rho=m\pi$  carry D0-brane charge $P=\frac{L^4 m^2}{16\alpha'^2}$. In turn, the NS five branes defined by  the dual partition of $N$ should be located at $\rho=\pi \sqrt{p(p+1)-m(m-1)}$, such that the relation  in eq.(\ref{D0NS5}) is satisfied. 
 
 Below, we will use all the information collected in this section to propose
a field theory dual to our non-Abelian T-dual solution of eq.(\ref{natdads})

\section{The non-Abelian T-dual as a BMN vacuum}\label{QFT}

We have seen in the previous section that our non-Abelian T-dual solution fits in the LLM classification of $SU(2|4)$ solutions \cite{Lin:2004nb}, with a potential given by eq. (\ref{vv}).  We have also seen that the solution is singular at $r=0$ (or $\sigma=1$). Both the metric and dilaton approach, close to the singularity, those of NS five branes, as shown in eq.(\ref{qqq}). Interestingly, this singularity occurs exactly at the point where the radial component of the electric field, 
derived from the potential of eq.(\ref{vv}), vanishes. Following our discussion in Section \ref{Laplace} this suggests that the NATD solution may come up as a result of  taking a `coarse-graining' approximation in the Lin-Maldacena distribution of conducting plates associated to the solution.

Indeed, using the expressions in eq. (\ref{QQ}), we can derive the charge density that approximates our distribution of conducting plates taking the partition of $N$ suggested by the quantised charges of the non-Abelian T-dual solution. This partition is given by eq.(\ref{breaking2}), as found out in Section  \ref{charges}. Doing this one finds that,
\begin{equation}
\label{charge}
Q(\eta)=\frac{L^4}{16\alpha'^2}\eta= \alpha \eta\, .
\end{equation}
Plugging now this expression for $Q(z)$ in the `coarse-grained' potential given by eq. (\ref{potentialcont}), and using that ---see \cite{Donos:2010va},
\begin{equation}
\label{potreg}
\int_0^\infty dz \frac{z}{\sqrt{\sigma^2+(z-\eta)^2}}=\lim_{L\rightarrow \infty}\Bigl[\frac12 \int_0^L dz\frac{2z}
{\sqrt{\sigma^2+(z-\eta)^2}} + L+\eta\log{2L}\Bigr]
\end{equation}
where, as explained in \cite{Donos:2010va}, the $\eta\log{2L}$ term can be added in order to regularise the potential, and not changing the expression $\dot{V}$ of the background, 
we are able to reproduce the potential associated to the NATD solution, given by eq.(\ref{natdads}).

The result of this analysis is that the NATD solution arises as a `coarse-grained' approximation of the Lin-Maldacena problem associated to it. This justifies the existence of the singularity at $r=0$ (or $\sigma=1$). However, an important aspect of the solution remains that obscures its interpretation as dual to a PWMM vacuum. As we have seen, the large $\sigma$ asymptotics of the solution
is that of D0-branes smeared on $\mathbb{R}^3(\eta,\Omega_2)$. Consistently with this, we have shown that it satisfies a smeared version of the fluctuated equations written by Lin in \cite{Lin:2004kw}. Our solution thus represents a smeared-deformation of the PWMM.


Following the ideas in \cite{Lozano:2016kum,Lozano:2016wrs}, we would like to propose that the configuration obtained using non-Abelian T-duality---eq.(\ref{natdads})-- represents a {\it patch}
of a background with the right D0-brane asymptotics of  a faithful dual representation of the PWMM.
Thus, in this `completed' background the large $\sigma$ and large $\eta$ effects of smearing the D0 branes, as explained in Sections \ref{asymptotics} and
\ref{lin}, are amended. The consequence of this will be to have a $(0+1)$ dual field theory, a matrix model.


A simple way to achieve this is to take the charge distribution given by eq. (\ref{charge}) but with $\eta$ living in a finite interval, $\eta\in [0,L]$, as suggested by the regularised potential in eq.(\ref{potreg}). Ending the density of $SU(2)$ representations in this manner has the effect of recovering the asymptotic behaviour driven by $V\sim \eta \sigma^2-\frac{2}{3}\eta^3$, and thus the large distance asymptotics of D0 branes.


The explicit potential generated by the charge distribution in eq. (\ref{charge}) with $\eta\in [0,L]$ is, using eq.(\ref{potentialcont}),
\begin{eqnarray}
\label{pppp}
& & V= (\eta\sigma^2 -\frac{2}{3}\eta^3)-2\alpha \eta\log\sigma+ \alpha \sqrt{\sigma^2+(L-\eta)^2} - \alpha     \sqrt{\sigma^2+(L + \eta)^2}  \\
& & +\alpha \eta\log\Big( \big[(L-\eta) +\sqrt{\sigma^2+(L-\eta)^2} \big] \big[(L+\eta) +\sqrt{\sigma^2+(L+\eta)^2} \big] \Big).\nonumber
\end{eqnarray}
One can check that this potential satisfies the Laplace equation (\ref{laplace}), and that for finite $L$ the asymptotics is that of the $(\eta\sigma^2 -\frac{2}{3}\eta^3)$ term. Taking instead $L$ very large
we find,
\begin{equation}
V\sim(\eta\sigma^2 -\frac{2}{3}\eta^3)-2\alpha \eta\log\sigma + \Delta V,
\label{mmmm}
\end{equation}
with 
\begin{equation}
\frac{1}{\alpha}\Delta V= -2\eta (1-\log 2-\log L) -\frac{3}{2L^2}(\eta\sigma^2-\frac{2}{3}\eta^3) +O(\frac{1}{L^4}).\nonumber
\end{equation}
The first term in $\Delta V$ does not contribute to $\dot{V}$ nor to the background in eqs.(\ref{LM})-(\ref{LM2}).
All subleading terms in the expansion satisfy the Laplace equation. Thus, taking $L\rightarrow \infty$ we encounter the $V_{\textrm{NATD}}$, which produces the solution in eq.(\ref{natdads}), whose large distance behaviour is that of smeared D0 branes. This can be
interpreted as the deformation of the dual PWMM with an irrelevant operator, which mirrors the ever-growing density of $SU(2)$ representations, $Q(z)=z$ in that system. This  is an effect of focusing on the solution given by $V_{\textrm{NATD}}$.

On the other hand, the full {\it completed} solution obtained with the potential in eq.(\ref{pppp})  replaces the ever-growing density of representations, by that generated with eqs.(\ref{potentialcont}) and (\ref{charge}) up to a maximum value of $\eta$, $\eta=L$. This  ameliorates the long distances behaviour, by using the background generated by the potential in eq.(\ref{pppp}). We are working with  a matrix model deformed by relevant operators. 
As anticipated in the papers \cite{Bak:2005ef}, \cite{Shieh:2007xn} and \cite{Donos:2010va}, this procedure does not resolve the 
small distance singularity. Indeed, what happens is that the singularity (originally at $\sigma=1$ for the non-Abelian T-dual solution), moves to $\sigma=0$ for the completed background obtained with the potential in eq.(\ref{pppp}). In our case, this singularity could be screened by a black hole horizon adding temperature to the dual PWMM.

In summary, the non-Abelian T-dual solution should be thought of as a patch of a  more general background. This {\it completed} solution, encoded by the potential in eq.(\ref{pppp}), is dual to a vacuum of  a $(0+1)$ quantum field theory.

 \section{Connection with the $AdS_7\times S^4$ superstar}\label{penrosesection}

In the previous section we have found that the geometry in eq.(\ref{natdads}) can be {\it completed}
using the ideas of \cite{Bak:2005ef,Shieh:2007xn,Donos:2010va}, in such a way that the asymptotics of the metric becomes that of D0-branes. Following this idea, the geometry of eqs.(\ref{natdads})-(\ref{xxy}), encoded by the potential function $V_{\textrm{NATD}}$ of eq.(\ref{vv}), was completed by replacing the potential with that found in eq.(\ref{pppp}).

This completion was motivated by the fact that the `smeared D0-branes' asymptotics
of the NATD solution was in conflict with its interpretation as dual to a vacuum of a 
$(0+1)$ dimensional QFT such as the PWMM. Even if the completion in Section \ref{QFT} gives rise to the right D0-brane asymptotics, a singularity, this time at $\sigma=0$, still remains, which indicates that some effects of the QFT are not captured by the completed supergravity solution.


In this section, we will propose a different way of globally defining our non-Abelian T-dual geometry by 'growing dimensions' to the $(0+1)$-QFT. More specifically, we will show 
that we can connect our background with the Penrose limit of the superstar solution in 
 \cite{Leblond:2001gn}, which is dual to a relevant deformation of the 6d CFT living in M5-branes.

We will start by studying in detail the Penrose limit of the $AdS_7\times S^4$ superstar solution constructed in  \cite{Leblond:2001gn}, that describes  giant gravitons backreacting
on $AdS_7\times S^4$. We will show that this geometry exactly agrees with our non-Abelian T-dual solution uplifted to eleven dimensions.
A possible interpretation of non-Abelian T-duals as Penrose limits of  AdS solutions was suggested in the seminal paper \cite{Sfetsos:2010uq} by Sfetsos and Thompson. Our explicit realisation of this idea --the first in the literature-- suggests that it may apply as well to other non-Abelian T-dual geometries.

   \subsection{Penrose limit of the half-BPS $AdS_7\times S^4$ superstar solution}\label{ssuperstar}
   
   As we have discussed before, by relaxing one of the boundary conditions of the Laplace equation we can still generate physically relevant solutions.
    An interesting example is the deformation by giant gravitons of $AdS_7 \times S^4$,
    known as the superstar solution \cite{Leblond:2001gn}. Such a deformation breaks half of the supersymmetries of the vacuum solution and corresponds to a horizon free black hole  
  whose singularity is interpreted as a condensate of giant gravitons. Interestingly, it was found that the origin of the singularity in the electrostatic description corresponds to the coarse grained limit of a distribution of conducting disks whose sizes are much bigger than the separation between them, such that far from the disks we can replace the discrete configuration by a conducting continuous surface of shape $R(\eta)$ \cite{Donos:2010va}. Such a description was shown to reproduce exactly the microscopic interpretation of the solution. Let us now proceed by describing the relationship between this solution and the non-Abelian T-dual solution 
studied in this work. For concreteness, we will only work with
the eleven-dimensional metric, though the configuration is completed by a $F_4$-field. 
   
The metric of the eleven dimensional $AdS_7\times S^4$ superstar solution is described by \cite{Leblond:2001gn},
 \begin{align}\label{superstar}
 ds^2=&\Delta^{1/3}\left(-f H^{-1}dt^2+f^{-1}dr^2+r^2 d\Omega_5^2+L^2\frac{d\theta^2}{4}\right)\nonumber\\
 &+\frac{\Delta^{-2/3} L^2}{4}\left(\sin^2\theta\, d\Omega_2^2+H\cos^2\theta(dx_{11}+\frac{2}{H L}dt)^2\right),
 \end{align}
where 
\begin{align}\label{functions}
\Delta=1+\frac{Q}{r^4}\sin^2\theta,\quad H=1+\frac{Q}{r^4},\quad f=1+\frac{H}{L^2}r^2,
\end{align}
and $d\Omega_5^2$ and $d\Omega_2^2$ are the line elements of the five and two spheres respectively.
 To proceed consider  the following scalings,
\be \label{scaling1}
r=L\tilde{r},\qquad t=L\tilde{t},\qquad Q=\tilde{q}L^4, \quad x_{_{11}}=\frac{4}{L^2}\tilde{x}_{_{11}}.
\e
{After these scalings, 
the metric in eq. (\ref{superstar}) reads,}
 \begin{align}\label{superstarr}
 ds^2=&L^2\left[\tilde{\Delta}^{1/3}\left(-\tilde{f} \tilde{H}^{-1}d\tilde{t}^2+
 \tilde{f}^{-1}d\tilde{r}^2+\tilde{r}^2 d\Omega_5^2+\frac{d\theta^2}{4}\right)\right.\nonumber\\
 &\left.+\frac{\tilde{\Delta}^{-2/3}}{4}\left(\sin^2\theta\, d\Omega_2^2+\frac{\tilde{H}\cos^2\theta}{L^4}(4d\tilde{x}_{11}+\frac{2 L^2}{\tilde{H}}d\tilde{t})^2\right)\right],
 \end{align}
{where,}
\begin{equation}\label{various}
\tilde{\Delta}=1+\frac{\tilde{q}}{\tilde{r}^4}\sin^2\theta,\qquad \tilde{H}=1+\frac{\tilde{q}}{\tilde{r}^4},\qquad \tilde{f}=1+\tilde{r}^2\tilde{H}.
\end{equation}
{We study the behaviour of the metric} 
in the neighbourhood of $\tilde{r}=0,\, \theta=0$. 
We take the limit $L\rightarrow\infty$ and let,
\begin{equation}\label{change1}
\tilde{r}=\frac{r_5}{L},\quad \theta=\frac{2}{L}r_2.\quad q=\tilde{q}L^2.
\end{equation}
Notice that we have rescaled the charge parameter in 
order to obtain a 
{ non trivial} $q$-dependent solution in the limit. 
The leading order behaviour of the functions in eq. \eqref{various} in the limit $L\rightarrow \infty$
is,
\begin{equation}
\tilde{\Delta}\sim 1+\frac{4qr_2^2}{r_5^4},\quad \tilde{H}\sim\frac{qL^2}{r_5^4},\quad \tilde{f}\sim 1+\frac{q}{r_5^2}.
\label{vvvbb}\end{equation}
The metric becomes,
\begin{equation}\label{lmetric}
ds^2=\tilde{\Delta}^{1/3}\left[-\left(1+\frac{q}{r_5^2}\right)\frac{r_5^4}{q}dt^2+\frac{dr_5^2}{1+\frac{q}{r_5^2}}+r_5^2d\Omega_5^2+dr_2^2+\frac{r_2^2}{\tilde{\Delta}}d\Omega_2^2+\frac{q}{4r_5^4\tilde{\Delta}}\left(4d\tilde{x}_{_{11}}+\frac{2r_5^4}{q}dt\right)^2\right].
\end{equation}

Redefining $r_5^2=4(\sigma^2-1),\, r_2=2\eta$ 
and setting $q=4$, the  metric in eq. \eqref{lmetric} can be written as,
\begin{equation}
\begin{split}
ds_{11}^2=&(\sigma^2-1)^{1/3}(4\eta^2+(\sigma^2-1)^2)^{1/3}\left(-4\sigma^2dt^2+\frac{4}{\sigma^2-1}(d\sigma^2+d\eta^2)+4d\Omega_5\right.\\
&+\left.\frac{4\eta^2(\sigma^2-1)}{4\eta^2+(\sigma^2-1)^2}d\Omega_2\right)+\frac{1}{(\sigma^2-1)^{2/3}(4\eta^2+(\sigma^2-1)^2)^{2/3}}\left(dx_{_{11}}-2(\sigma^2-1)^2dt\right)^2,
\end{split}
\end{equation}
{In summary, by focussing (or Taylor expanding) 
the solution in eq. (\ref{superstarr}) 
near the singularity at $ \tilde{r}=0$ 
along $\theta=0$, with the appropriate scalings in 
eq. (\ref{change1}), we find the eleven dimensional 
uplift of the 
non-Abelian T-dual solution in eq. (\ref{xxy}). 
Since Taylor expanding a solution does not (in general) 
produce a new solution of the equations of motion it 
is worth to investigate 
the nature of the above limit. 
In the following we will show that the procedure discussed 
above corresponds to a Penrose limit}.

{Consider a null geodesic $\gamma$ parametrised 
by the coordinates $(t,x_{11})$ that sits 
at the point $\theta=0, \tilde{r}= 0$ of the metric in eq.(\ref{superstarr}). 
We approach this point by taking 
$\tilde{r}=r_0+ \frac{r_5}{L}$ and $\theta=\theta_0+ \frac{r_2}{2L}$. We will be
interested in the case of $r_0$ and $\theta_0$ vanishing, 
and  $L \rightarrow \infty$.}

{Before formally taking the 
limit of large $L$, we consider the induced metric for this geodesic. 
It reads,}

\begin{equation}\label{induced2}
ds_{\gamma}^2=\tilde{\Delta}_{0}^{1/3}
\left(-\frac{\tilde{f}_{0} L^2}{\tilde{H}_0}d\tilde{t}^2
+\frac{1}{4}\frac{\tilde{H}_{0} \cos^2\theta_0}{
L^2 \tilde{\Delta}_{0}}(4d\tilde{x}_{11}+2\frac{ L^2}{\tilde{H_0}}d\tilde{t})^2
\right),
\end{equation}
{where}
\begin{equation}
\tilde{\Delta}_{0}=1+\frac{\tilde{q}}{r_0^2}\frac{\sin^2\theta_0}{r_0^2}, 
\quad \tilde{f}_0=1+ r_0^2+\frac{\tilde{q}}{r_0^2},
\quad \tilde{H}_0=1 +\frac{\tilde{q}}{r_0^4}.
\end{equation}
In the variable $dz= 4 dx_{11} + \frac{2 L^2}{\tilde{H}_0} dt$, we find the equation for the null geodesic,
\begin{equation}
\left(\frac{dt}{dz}\right)^2= \frac{\tilde{H}_0^2 \cos^2\theta_0}{4 L^2 \tilde{\Delta}_0 \tilde{f}_0}.\label{lalazzz}
\end{equation}

{Notice that in the limit $r_0\rightarrow 0$, 
the induced metric in eq. (\ref{induced2}) and the geodesic solution in eq.(\ref{lalazzz}) are ill defined as 
$H_0\rightarrow \infty$. 
This is expected since we are  considering a geodesic in the neighbourhood 
of a singularity. 
However, there is a nice conspiracy in 
the limit $L\rightarrow\infty$ such that 
the induced metric in eq. (\ref{induced2}) is well defined\footnote{
Using a geodesic that sits on the singularity $\tilde{r}=0$ is justified
by the fact that, according to the criteria presented in
  \cite{Maldacena:2000mw}, the singularity is 'acceptable/good'. 
There are various examples in the bibliography, in which probes placed at
'good' singularities calculate observables in a correct way.}. 
Indeed by taking the limit of large $L$ }
{and choosing null coordinates such that }
\begin{equation}
dt=\frac{\tilde{\Delta}_0^{1/6}}{2\sqrt{\tilde{f}_0} \cos\theta_0}du,
\quad dx_{11}=\frac{L^2\tilde{\Delta}_0^{2/3}}{4 \tilde{H}_0 \cos\theta_0}
\left(\frac{1}{\cos\theta_0}-\frac{1}{\sqrt{\tilde{\Delta}_0 \tilde{f}_0}}\right)du+dv,
\end{equation}
{in the limit in which $r_0\to 0$ and $\theta_0\to 0$ with  $L\rightarrow \infty$ and also scaling 
$ \tilde{q}=\frac{q}{L^2}$--- the content of eq.(\ref{change1}), 
the metric in eq.  (\ref{superstarr}) takes the form of a 
pp-wave in Rosen coordinates. Namely}
\begin{equation}\label{sol}
ds^2=2 du d v+\tilde{\Delta}^{1/3}\left(\frac{dr_5^2}{\tilde{f}}+r_5^2 d\Omega_5^2\right)+\tilde{\Delta}^{-2/3}\left(\tilde{\Delta} dr^2_2+r_2^2 d\Omega_2^2 \right),
\end{equation}
where $\tilde{\Delta},\tilde{f}$ are defined in eq.(\ref{vvvbb})
{We therefore see that the non-Abelian T-dual 
background corresponds to the Penrose limit of the 
half-BPS $AdS_7\times S^4$ superstar. 
Notice that the non-trivial rescaling in the charge parameter 
in eq. (\ref{change1}), is 
necessary in order to keep the effect of the 
parameter $\tilde{q}$ that controls a relevant deformation 
on the $(0,2)$ CFT. Had we rescaled the $\tilde{q}$ parameter 
in a different way such that the limit $L\rightarrow \infty$ 
remains finite we would have obtained the maximally supersymmetric 
pp-wave in eleven dimensions. 
Indeed, setting  $\tilde{q}=0$ in the metric of 
eq. (\ref{sol}), we obtain the maximally supersymmetric pp-wave, as expected.
}

{To make our argument stronger,
 notice that} the null geodesic chosen in eq. (\ref{change1}) 
is consistent (up-to typos) with the analysis performed 
in \cite{Alishahiha:2005nq},  
where the Penrose limit of the superstar solution 
in eq.(\ref{superstar}) was studied in the context of the 
LLM formalism. In this approach the Penrose limit 
is taken by blowing up the neighbourhood around 
the edge of a droplet describing the superstar solution. 
The upshot of this procedure is the potential function 
in eq. (\ref{vv}) which corresponds to the non-Abelian T-dual solution.

In summary, the Penrose limit of the $AdS_7\times S^4$ superstar coincides with the uplift to eleven dimensional supergravity of the  geometry in eq.(\ref{xxy}). 
Using Holography, this tells us that the CFT dual to our background in eq.(\ref{natdads}) could be thought of as a particular sector of the $(0,2)$ SCFT with relevant deformations. 


\section{Connection with the Sfetsos-Thompson solution}\label{connectionst}

In this section we show that our non-Abelian T-dual solution (obtained upon dualisation on an $S^3\subset AdS_5$) can also be obtained with  a Wick rotation on the Sfetsos-Thompson solution  \cite{Sfetsos:2010uq}  (obtained upon dualisation on an $S^3\subset S^5$). This observation, combined with the previous interpretation of our non-Abelian T-dual solution as a Penrose limit, suggests that it should be possible to interpret the Sfetsos-Thompson solution as the result of taking a Penrose limit on a globally well-defined $\mathcal{N}=2$ geometry.

Indeed, the Sfetsos-Thompson solution is an example of an  $AdS_5$ background belonging to the Gaiotto-Maldacena class of geometries  \cite{Gaiotto:2009gz}. Inspired by this, an interpretation to its dual CFT could be given in terms of an infinitely long linear quiver   \cite{Lozano:2016kum}.  Starting with the $AdS_5\times S^5$ solution,

\begin{align}
ds^2=&L^2 ds_{AdS_5}^2+ L^2 \Big[4 d\alpha^2
+4 \sin^2\alpha d\beta^2 +  \cos^2\alpha (\omega_1^2+\omega_2^2+\omega_3^2) \Big],\nonumber\\
& \qquad\qquad\qquad F_5= (1+ \star) \frac{64 }{g_sL^4} R^3 dR \wedge  d^4 x  \label{ads5xs5}, 
\end{align}
where $4 L^4= \pi g_s N \alpha'^2$ and $\omega_i$ are the $SU(2)$ left-invariant Maurer-Cartan forms,   
 the Sfetsos-Thompson solution was obtained 
acting with non-Abelian T-duality on one of the two $SU(2)$ isometries of the internal space. The resulting background is given by  \cite{Sfetsos:2010uq},
\begin{align}
\widehat{ds^2}=&4L^2ds_{AdS_5}^2
+ L^2 \Big[4d\alpha^2
+4\sin^2\alpha d\beta^2 \Big]+\frac{\alpha'^2}{L^2\cos^2\alpha} 
d\rho^2 +\frac{\alpha'^2L^2\cos^2\alpha \rho^2}{\alpha'^2 \rho^2 
+ L^4\cos^4\alpha}(d\chi^2 +\sin^2\chi d\xi^2),\nonumber\\
& \widehat{B}_2=\frac{\alpha'^3 \rho^3}{\alpha'^2 \rho^2 +L^4 \cos^4\alpha}
\sin\chi d\chi \wedge d\xi,\;\;\; 
e^{-2\widehat{\Phi}}=\frac{ L^2 \cos^2\alpha}{\alpha'^3}
(L^4 \cos^4\alpha +\alpha'^2 \rho^2),\nonumber\\
& \widehat{F}_2=\frac{8L^4    }{\alpha'^{3/2}}
\sin\alpha \cos^3\alpha d\alpha \wedge d\beta,\;\;  \hat{A}_1=-\frac{2L^4   }{\alpha'^{3/2}} \cos^4\alpha d\beta. \;\; \widehat{F}_4= \widehat{B}_2\wedge \widehat{F}_2.
\label{ads5xs5natd}
\end{align}
A potential function satisfying the Laplace equation (\ref{laplace}) can be written, in this case with a charge density given by $\lambda(\eta)=\sigma\partial_{\sigma}V(\sigma,\eta)|_{\sigma=0}$.  After changing variables as,
\begin{equation}
\rho=\frac{2L^2}{\alpha'}\eta,\quad \sigma=\sin\alpha,
\end{equation}
the potential is
\begin{equation}\label{vst}
V_{ST}=\eta(\log\sigma-\sigma^2)+\frac{\eta^{3}}{3}.
\end{equation}

Let us now Wick rotate the solution in eq. (\ref{ads5xs5natd}), where
\begin{equation}
ds_{AdS_5}^2=-\cosh^2 \tilde{r} dt^2+d\tilde{r}^2+\sinh^2\tilde{r} d\Omega_3^2\, ,
\end{equation}
in global coordinates, by performing the following analytical continuation in the coordinates: $\alpha\rightarrow  -i r-\frac{\pi}{2}$, $t\rightarrow \phi$ and $\tilde{r}\rightarrow i\theta$. The effect of the above transformations is to replace ,
\begin{equation}
 \cos\alpha\rightarrow -i\sinh r,\quad ds^2_{AdS_5}\rightarrow -d\Omega_5^2,
\end{equation}
in eq. (\ref{ads5xs5natd}), where $d\Omega_5^2$ is given by,
\begin{equation}
d\Omega_5^2=\cos^2{\theta} d\phi^2+d\theta^2+\sin^2\theta d\Omega_3^2.
\end{equation}
In order to get the metric with the correct signature we further impose $L\rightarrow 2i \tilde{L}$. 
In this way one arrives at the non-Abelian T-dual  solution in eq. (\ref{xxy}). Notice that the relation between the solutions in eqs. (\ref{xxy}) and  (\ref{ads5xs5natd}) is an explicit ten dimensional realisation of the
analytical continuation procedure discussed in \cite{Lin:2004nb}.

We would like to stress that even if both solutions in eqs. (\ref{xxy}) and  (\ref{ads5xs5natd}) solve the same Laplace equation (\ref{laplace}), they differ in the boundary conditions they satisfy. Notice as well that it is necessary to change the sign of the potential, $V_{ST}\rightarrow -V_{NATD}$, to ensure that the metric gets the correct signature.




\section{Conclusions and future directions}\label{conclusiones}
For the benefit of the reader, let us start with a  brief summary of this work.

In this paper, we constructed  and studied a new background, solution to the equations of motion of Type IIA supergravity. It was obtained by the application of non-Abelian T-duality on an $SU(2)$ inside the $SO(2,4)$ isometry
of $AdS_5\times S^5$. In Section \ref{sectionNATD}, we worked-out  the asymptotics, conserved charges and interesting probes of the space-time. A proposal for a quantum field theory dual  in terms of a given BMN vacuum was then given in Section \ref{QFT}. The precise way in which we were able to define a consistent field theory dual associated to our solution provided a very explicit realisation of the idea that  solutions generated by non-Abelian T-duality correspond to a patch of a more generic background. The  quantum field theory proposed in Section \ref{QFT} to inform the completion of our Type IIA background is $(0+1)$-dimensional, a matrix model obtained by a relevant deformation of the BMN Plane Wave Matrix Model. As we emphasised the background is still singular at short distances ($\sigma=0$), being this a feature of the procedure of \cite{Bak:2005ef},  that we adopted.  Meanwhile, in Section \ref{penrosesection} a different completion of the non-Abelian T-dual background and its associated field theory, was proposed in terms of a 6d (0,2) SCFT, also deformed by relevant operators. The material in Section \ref{penrosesection}
explicitly realises (by the first time) a conjecture that appeared in various previous instances in the literature. Namely, our new background is the Penrose limit of a well-known solution, the Superstar in $AdS_7\times S^4$.
Finally, in Section \ref{connectionst} we connected our Type IIA background with that constructed in \cite{Sfetsos:2010uq}.

The relevance of the material we have presented immediately suggests numerous new avenues of research. Let us mention some of them here.

It would be interesting to extend the realisation of the non-Abelian T-dual solution as the Penrose limit of the $AdS_7\times S^4$ solution to other examples. It would also be 
nice to study more exhaustively the relation between our non-Abelian T-dual solution and the Sfetsos-Thompson solution presented in Section \ref{connectionst}. This would amount to a way of connecting Gaiotto-Maldacena backgrounds (dual to Gaiotto field theories) with Lin-Maldacena solutions (dual to $SU(2|4)$ quantum field theories). 

It is clear that---by construction-- the completion proposed in Section \ref{QFT} to get a well-defined $(0+1)$ dual QFT is just one among many possible ones. The only reason we completed in the way presented here is to ease the calculations. But this in turn, suggests that there exist a host of different quantum field theories and associated string duals that share a common sector, represented by the solution obtained via non-Abelian T-duality. It would be interesting to study this universality
in more detail, which points to a generalised form of Eguchi-Kawai mechanism of reduction \cite{Ishii:2008ib}. 

It would also be of interest to combine our solution and its completion with the black-hole geometry discussed in Section \ref{sectionNATD}. The cloaking of the singularity makes it a perfect arena to study finite temperature effects. A computer simulation of the matrix model can be compared  with the  predictions calculated from our Type IIA background supplemented by the black hole. Indeed, following the lead of the papers
 \cite{Berkowitz:2016tyy}, a simulation for the PWMM at finite temperature
should be feasible. Then, correlators calculated with the string background
could be checked against lattice simulations.

There are also a number of points of more technical nature that it would be nice to clarify. One of them is the study of the multipole expansion that follows from sub-leading terms to eq.(\ref{potpwmm}), and the similar $1/L$-sub-leading terms of eq.(\ref{mmmm}). The study of the different terms in the expansion, might contain interesting structure. It would also be of interest to study some sectors of the (0,2) SCFT in six dimensions in terms of the PWMM. Similarly, the fact that our completion in Section \ref{penrosesection} involves M5-branes points to a feature  that threads many different solutions obtained via non-Abelian T-duality. We hope to report on these and other interesting questions soon.

\section*{Acknowledgments:} 
In conversations and sharing their views, many physicists have helped us to improve the presentation of the material of this paper. We are thankful to
Adi Armoni, S. Prem Kumar, Horatiu Nastase, Diego Rodriguez-Gomez and Daniel Thompson.
Y.L. is partially supported by the Spanish and Regional Government Research Grants FPA2015-63667-P and FC-15-GRUPIN-14-108. Y.L. would like to thank the Physics Department of Swansea U. for the warm hospitality during the completion of this work.
C. N is Wolfson Fellow of the Royal Society. S. Z. is a Newton International Fellow of the Royal Society. 





\end{document}